%% file: main.tex
\documentclass[a4paper]{JHEP3}

\usepackage{amsmath}
\usepackage{cite}
\usepackage{multirow}
\usepackage{array}
\usepackage{feynarts}

\newcommand{\bea}{\begin{eqnarray}}

\newcommand{\eea}{\end{eqnarray}}

\newcommand{\bei}{\begin{itemize}}

\newcommand{\eei}{\end{itemize}}

\newcommand{\bean}{\begin{eqnarray*}}

\newcommand{\eean}{\end{eqnarray*}}
           
\newcommand{\nn}{\nonumber \\}

\newcommand{\HPL}{\text{H}}

\def\eps{\epsilon}

\def\magnus #1{\Omega [ #1]}
\def\ef #1{f_{#1}}
\def\top #1{\mathcal{T}_{#1}}

\allowdisplaybreaks

\title{
Magnus and Dyson Series for Master Integrals
}

\author{Mario Argeri \\ 
Dipartimento di Scienze e Innovazione Tecnologica, Universit\`a del
Piemonte Orientale, Viale Teresa Michel 11, I-15121 Alessandria, Italy \\
E-mail: \email{mario.argeri@mfn.unipmn.it}
}

\author{Stefano Di Vita \\ 
Max-Planck-Institut f\"ur Physik, F\"ohringer Ring 6, D-80805 M\"unchen, Germany \\
E-mail: \email{divita@mpp.mpg.de}
}    

\author{Pierpaolo  Mastrolia \\ 
Max-Planck-Institut f\"ur Physik, F\"ohringer Ring 6, D-80805 M\"unchen, Germany \\
Dipartimento di Fisica e Astronomia, Universit\`a di Padova, and INFN Sezione di Padova, via Marzolo 8, 35131 Padova, Italy \\
E-mail: \email{pierpaolo.mastrolia@cern.ch}
}    

\author{Edoardo Mirabella \\ 
Max-Planck-Institut f\"ur Physik, F\"ohringer Ring 6, D-80805 M\"unchen, Germany \\
E-mail: \email{mirabell@mpp.mpg.de}
}

\author{Johannes Schlenk\\ 
Max-Planck-Institut f\"ur Physik, F\"ohringer Ring 6, D-80805 M\"unchen, Germany \\
E-mail: \email{jschlenk@mpp.mpg.de}
}    

\author{Ulrich Schubert \\ 
Max-Planck-Institut f\"ur Physik, F\"ohringer Ring 6, D-80805 M\"unchen, Germany \\
E-mail: \email{schubert@mpp.mpg.de}
}    

\author{Lorenzo Tancredi \\ 
Physik-Institut, Universit\"at Z\"urich,
Wintherturerstrasse 190, CH-8057 Z\"urich,
Switzerland\\ 
E-mail: \email{tancredi@physik.uzh.ch}
}    

\preprint{ MPP-2014-7, \; ZU-TH 01/14}

\abstract{ 
We elaborate on the method of differential equations for evaluating Feynman integrals. 
We focus on  systems of equations for master integrals  having a linear dependence on the dimensional
parameter. For these systems  we identify the criteria to bring them  in a canonical form, recently
identified by Henn, where the dependence of the dimensional
parameter is disentangled from the kinematics. 
The determination of the transformation and the computation of the solution are 
obtained by using  Magnus and Dyson  series expansion.
We apply the method  to planar and non-planar two-loop
QED vertex diagrams for massive fermions, and to non-planar two-loop
integrals contributing to $2 \to 2$ scattering of massless particles.
The extension to systems which are polynomial in the dimensional
parameter is discussed as well.
}

\begin{document}

\input{intro}

\input{quantum}

\input{magnusdyson}

\input{diffeq}

\input{bhabha1L}

\input{vertex2L}

\input{xbox2L}

\input{polyeps}

\input{conclusions}

\section*{Acknowledgments}
We wish to thank Roberto Bonciani for interesting discussions and comments on the manuscript.
The work of  P.M. and U.S. was supported by the Alexander von Humboldt
Foundation, in the framework of the Sofja Kovalevskaja Award Project ``Mathematical Methods for Particle Physics'', endowed by the German Federal Ministry of Education and Research. 
The work of L.T. was supported in part by the Swiss National Science
Foundation (SNF) under  contract PDFMP2-135101. M.A.  wishes to acknowledge the kind hospitality of the Max Planck Institut f\"ur Physik 
in Munich  during the completion of  this project. The Feynman diagrams depicted in this
paper are drawn using {\sc FeynArts}~\cite{Hahn:2000kx}.

\appendix
\input{appendix}

\input{MIvertex}

\input{MIxbox}

\bibliographystyle{JHEP}

\bibliography{references}

\end{document}

%% file: intro.tex
\section{Introduction}
\label{sec:intro}

The {\it method of differential equations} (DE's), developed by
Kotikov, Remiddi and Gehrmann ~\cite{Kotikov:1990kg,Remiddi:1997ny,Gehrmann:1999as} and reviewed in Ref.~\cite{Argeri:2007up,Smirnov:2012gma}, is one of the most effective
techniques for computing dimensionally regulated multi-loop integrals and has led to significant achievements in the
context of multi-loop corrections. Within the continuous dimensional
regularization scheme, Feynman integrals can be related by using integration-by-parts identities
(IBP-id's)~\cite{Chetyrkin:1980pr,Tkachov:1981wb}, Lorentz invariance
identities~\cite{Gehrmann:1999as}, Gram
identities~\cite{Gluza:2010ws}, and quasi-Shouten
identities~\cite{Remiddi:2013joa}.   These relations can be exploited
in order to {\it identify} a set of independent integrals, dubbed {\it
  master integrals} (MI's),  that can be used as a basis of functions for the 
virtual contributions to scattering amplitudes. 

The MI's are functions of the kinematic invariants
constructed from the external momenta, of
the masses of the external  particles and  of the
particles running  in the loops, as well as of the number of spacetime dimensions.
Remarkably,  the existence of the aforementioned  relations forces the MI's
to obey linear systems of first-order differential equations in
the kinematic invariants,
which can be used for the determination of their expression. 
When possible, these systems are solved exactly for generic values of the space-time dimension  $D$. Alternatively, they can be Laurent-expanded around suitable values of the dimensional parameter up to the required order, 
obtaining a system of chained differential equations for the
coefficients of the expansions.  
In the most general case, the latter are finally  integrated by using the method of  Euler's variation of 
constants. 

The nested structure of the Laurent expansion of the
linear system  leads to an iterative structure for the solution that, order-by-order in $\epsilon = (4-D)/2$, is
written in terms of repeated integrals, starting from the  kernels dictated by the homogeneous solution. The 
transcendentality of the solution is  associated to the number of repeated integrations and increases by one unit as 
the order of the $\epsilon$-expansion increases. 
{\it The} solution of the system, namely the MI's, is finally determined by imposing the {\it
  boundary conditions} at special values of the kinematic
variables,  properly chosen either in correspondence of configurations that reduce the MI's to   
 simpler integrals or in correspondence of pseudo-thresholds. In this
 latter case, the boundary conditions
 are obtained  by imposing  the {\it regularity} of the MI's  around  unphysical
singularities, ruling out divergent behavior of the general solution of the systems.

For any given scattering process the set of MI's is not unique, and, in practice, 
their choice is rather arbitrary. Usually MI's are identified 
after applying the Laporta reduction algorithm~\cite{Laporta:2001dd}.
Afterward, convenient manipulations of the basis of MI's may be performed.

Proper choices of MI's can simplify the form of the 
systems of differential equations and, hence, of their solution, although general criteria for
determining such optimal sets are not available.
An important step in this direction has been recently taken in
Ref.~\cite{Henn:2013pwa}, where Henn proposes to solve the systems of DE's
for MI's with algebraic methods. The key observation 
 is that a {\it good} choice of MI's allows one to cast  the system of DE's in a {\it canonical
form}, where the dependence on $\epsilon$, is factorized from the
kinematic. 
The integration of a system in canonical form trivializes and the analytic properties of its general  solution are manifestly
inherited  from the matrix associated to the
system, which is  the kernel of the
representation of the solutions in terms of repeated integrations.

This novel idea has  been applied in a number of cases by Henn,
Smirnov, and Smirnov \cite{Henn:2013tua,Henn:2013woa,Henn:2013nsa},
showing the effectiveness of this approach.
As pointed out in \cite{Henn:2013pwa}, finding an algorithmic  procedure which,
starting from a generic set of MI's, leads to a set MI's fulfilling a
canonical system of DE's is a formidable task. In practice, the
quest for the suitable basis of MI's is determined by qualitative
properties required for the solution, such as finiteness in the $\epsilon \to 0$
limit, and homogeneous transcendentality, which turn into quantitative
tools like the unit leading singularity criterion and the {\it dlog}
representation in terms of Feynman parameters \cite{Gehrmann:2011xn}.

\bigskip
In this article, we suggest a convenient form  for the
initial system of MI's, and we  propose an algorithm to find
the transformation matrix yielding  a canonical system. 
In particular, we choose a set of MI's obeying to a system of
DE's which has a {\it linear} $\epsilon$-dependence,
and we find a transformation which absorbs the $\mathcal{O}(\epsilon^0)$
term and leads to a new system of DE's where the $\epsilon$-dependence
is factorized. This transformation, as well as the integration of the 
canonical system, are obtained by using  Magnus and Dyson series
expansions \cite{Magnus,Dyson:1949bp,Blanes:arXiv0810.5488}.
The  procedure we propose can be generalized to the case of
systems that are {\it polynomial} in $\epsilon$. Nevertheless, for the
cases at hand, we have succeeded to begin from a set of
MI's obeying a system that is linear  in $\epsilon$.
We show the effectiveness of our method by applying it to
non-trivial integrals. In particular,  we apply our procedure to  determine the MI's of 
 the two-loop vertex diagrams contributing to the massive fermion
form-factors in QED~\cite{Bonciani:2003te,Bonciani:2003ai} and of  the non-planar two-loop diagrams
contributing to the $2 \to 2$ scattering of massless particles~\cite{Tausk:1999vh,Anastasiou:2000mf}.
Together with the ones in Ref.~\cite{Henn:2013woa}, the set of MI's for the two-loop QED vertices hereby presented
constitute a transcendentally-homogeneous subset for tackling the
analytic calculation of the still unknown non-planar two-loop box diagrams contributing
to the  massive Bhabha scattering in QED~\cite{Czakon:2004wm,Bonciani:2005im,Czakon:2006pa}.  It may enter  as well in  more general classes
of scattering processes involving massive particles.

Let us finally remark, that, while the canonical form of the system guaranties an easy integration procedure, 
it alone does not directly imply the homogeneous transcendentality of the solution. 
Indeed, this property may be affected by the analytic properties of additional
inputs such as the boundary conditions and the integrals which appear in the system of
DE's but cannot be determined from it. The latter are   integrals whose differential equation is
homogeneous and carries only the scaling information.

The  paper is structured as follows.
In section 2, we  present the Quantum Mechanical example that inspired this
study. The definition of Magnus series and its connection to the Dyson series are presented  in Section 3. 
In Section 4 we show how to derive the canonical system starting from a 
linear $\epsilon$-dependent system.
In Section 5, 6 and 7,  we apply our procedure to 
the one-loop massive Bhabha scattering in QED,
to the two-loop vertex diagrams contributing to the massive electron
form factors in QED,
and to the two-loop non-planar box diagram respectively.
In Section 8, we show how our method generalizes to the case of a system
of differential equations which is polynomial in $\eps$.
The properties of the matrix exponential and 
the proof of Magnus theorem are shown in Appendix~A, 
while the Appendices~B, and~C collect the expressions of the MI's 
of  the two-loop QED vertex diagrams and of  the two-loop non-planar 
box diagram respectively.

We used the computer code {\sc Reduze2}~\cite{Studerus:2009ye,vonManteuffel:2012np} for the generation of the
systems of differential equations.

This manuscript is accompanied by two ancillary files, containing the
results of the canonical MI's for the two-loop QED vertices, and for
the two-loop non-planar box, respectively.

%% file: quantum.tex
\section{On time-dependent perturbation theory}
Given an Hamiltonian operator $H$, we consider the Schr\"odinger equation ($\partial_t \equiv \partial / \partial t$)
\bea
i\hbar \ \partial_t | \Psi(t) \rangle =   H(t)   | \Psi(t) \rangle  \, .
\eea
Let us assume that ${H}$ can be split in two terms as 
\bea
{ H(t)}  = { H} _0(t) + \epsilon { H}_1(t) \ ,
\label{eq:Ham}
\eea
where  ${H} _0$ is a solvable Hamiltonian and $\epsilon \ll 1$ is a
small perturbation parameter. We may move to the  {\it interaction picture}  
by performing a  transformation via a unitary  operator  $B$. In this 
representation  any operator $A$ transforms according to
\bea
A(t) = B(t) A_I(t) B^\dagger(t) \; .
\eea
In the interaction picture one imposes that only $H_1$ ($H_0$)  enters the time evolution of the states (of the operators),  thus $B$ is obtained by
imposing
\bea
i \hbar \ \partial_t U_I(t)= \epsilon \, {H}_{1,I}(t) U_I(t)   +\left ( H_{0,I}(t) -i \hbar \,  B^\dagger(t) \ \partial_t B(t) \right ) U_I(t) \stackrel{!}{=}\epsilon \, {H}_{1,I}(t) U_I(t),  \;
\eea
so that  $B$ fulfills 
\bea
 i \hbar \,  \partial_t B(t) =H_{0}(t) B(t)\ .
 \label{Eq:eqB}
\eea
In the interaction picture the  Schr\"odinger equation  can be cast in a {\it canonical form},
\bea
i\hbar \ \partial_t | \Psi_I(t) \rangle = \epsilon \,   H_{1,I}(t)   | \Psi_I(t) \rangle \, ,
\eea
where the $\epsilon$-dependence is factorized.
If the Hamiltonian $H_0$ at different times commute, the solution of Eq.~(\ref{Eq:eqB}) is
\bea
B(t) = e^{- {i \over \hbar} \int_{t_0}^t d\tau H_0(\tau)}\ .
\eea

The important remark in this derivation is that, as a consequence of
the linear $\epsilon$-dependence of the original Hamiltonian Eq.~(\ref{eq:Ham}),
the states fulfill  an equation in a canonical form  
by means of a transformation matrix $B$ that obeys the differential
equation~(\ref{Eq:eqB}). This simple quantum mechanical example contains the 
two main guiding principles for building  canonical systems of differential equations for Feynman integrals:
\begin{itemize}
\item choose a set of Master Integrals obeying a system of
  differential equations linear in $\epsilon$;
\item find the transformation matrix by solving a differential
  equation governed by the constant term. 
\end{itemize}
In this example $H_0(t)$ and $B(t)$ commute.
In the case of Feynman integrals, no assumption can be made on the
properties of the matrix associated to the systems of DE's built out of IBP-id's. Therefore, 
in the following, we need to consider the generic case of non-commutative operators.

%% file: magnusdyson.tex
\section{Magnus series expansion}

Consider a generic linear matrix differential equation~\cite{Blanes:arXiv0810.5488}
\bea
\partial_x Y(x) = A(x) Y(x) \ , \quad Y(x_0) = Y_0 \ .
\label{Eq:mainSY}
\eea
 If  $A(x)$ commutes with its integral $\int_{x_0}^x d\tau A(\tau)$, {\it e.g.}  in the scalar case,
 the solution can be written as 
\bea
Y(x) = e^{\int_{x_0}^x d\tau A(\tau)} \ Y_0 \ .
\label{eq:solMagn0}
\eea
In the general non-commutative case, one can use the {\it Magnus theorem}
\cite{Magnus} to write the solution as,
\bea
Y(x) = e^{\Omega(x,x_0)} \ Y(x_0) \equiv  e^{\Omega(x)} \ Y_0 \,  ,
\label{eq:solMagn1}
\eea
where $\Omega(x)$ is written as a  series expansion,  called {\it Magnus expansion},
\bea
\Omega(x)=\sum_{n=1}^\infty \Omega_n(x) \ .
\label{eq:Magnus}
\eea
The proof of the Magnus theorem is presented in the Appendix~\ref{App:Magnus}, together with 
the actual expression of the  terms $\Omega_n$.
The first three terms of the expansion~(\ref{eq:Magnus}) read as follows:
\bea
\Omega_1(x) &=& \int_{x_0}^x d\tau_1 A(\tau_1)\  ,\nonumber  \\
\Omega_2(x) &=& 
{1 \over 2}
\int_{x_0}^x d\tau_1  
\int_{x_0}^{\tau_1} d\tau_2  \ 
[A(\tau_1), A(\tau_2)] \ ,\nonumber  \\
\Omega_3(x) &=& 
{1 \over 6}
\int_{x_0}^t d\tau_1  
\int_{x_0}^{\tau_1} d\tau_2  
\int_{x_0}^{\tau_2} d\tau_3  \ 
[A(\tau_1),[A(\tau_2),A(\tau_3)]] +
[A(\tau_3),[A(\tau_2),A(\tau_1)]] 
\label{eq:Magnus:exp} \ . \qquad
\eea
We remark that  if  $A$  and its integral commute, the series~(\ref{eq:Magnus})
is truncated at the first order,  $\Omega = \Omega_1$, and we recover 
the solution~(\ref{eq:solMagn0}).
As a notational aside, in the following we
will use the symbol
$\magnus{A}(x)$ to denote the Magnus expansion obtained using  $A$ as kernel.

\subsection{Magnus and Dyson series expansion}

Magnus series is related to the Dyson series~\cite{Blanes:arXiv0810.5488},
and their connection  can be obtained starting from the Dyson
expansion of the solution of the system~(\ref{Eq:mainSY}),
\bea
Y(x) = Y_0 + \sum_{n=1}^\infty Y_n(x) \ , \quad Y_n(x) \equiv \int_{x_0}^x d\tau_1 \ldots \int_{x_0}^{\tau_{n-1}} d\tau_n \ 
A(\tau_1) A(\tau_2) \cdots A(\tau_n) \ ,
\label{eq:Dyson}
\label{eq:Dyson:exp}
\eea
in terms of the  {\it time-ordered} integrals $Y_n$. 
Comparing Eq.~(\ref{eq:solMagn1}) and~(\ref{eq:Dyson}) we have 
\bea
\sum_{j=1}^\infty \Omega_j(x) = {\rm log} 
\left( Y_0 + \sum_{n=1}^\infty Y_n(x) \right) \ ,
\eea
and the following relations
\bea
Y_1 &=& \Omega_1 \ , \nonumber \\
Y_2 &=& \Omega_2 + {1 \over 2 !} \Omega_1^2 \ , \nonumber  \\ 
Y_3 &=& \Omega_3 + {1 \over 2 !} 
(\Omega_1 \Omega_2 +
\Omega_2 \Omega_1)
+ {1 \over 3 !} \Omega_1^3 \ ,  \nonumber \\ [0.5ex]
\vdots & & \qquad \vdots \nonumber \\[0.5ex]
Y_n &=& \Omega_n + \sum_{j=2}^n {1 \over j} Q_n^{(j)} \, .
\eea
The matrices $Q_n^{(j)}$ are defined as
\bea
Q_n^{(j)} = \sum_{m=1}^{n-j+1} Q_m^{(1)} Q_{n-m}^{(j-1)} \ ,
\quad 
Q_n^{(1)} \equiv \Omega_n \ , \quad  
Q_n^{(n)} \equiv \Omega_1^n \ . 
\eea
In the following, we will use both Magnus and Dyson series. 
The former allows us to easily demonstrate how a system of DE's, 
whose matrix is linear in $\epsilon$, can be cast in the canonical form.
The latter can be more conveniently used for the explicit
representation of the solution.

%% file: diffeq.tex
\section{Differential equations for Master Integrals}
\label{Sec:diffeq}
We consider  a linear system of first order differential equations
\bea
\partial_x f(\epsilon,x) =  A(\epsilon, x) \ f(\epsilon,x)\ ,  
\label{eq:MIdiffeq}
\eea
where  $f$ is a vector of MI's, while $x$ is a variable depending on kinematic invariants and masses.
We suppose that $A$ depends linearly on $\epsilon$,
\bea
A(\epsilon,x) = A_0(x) + \epsilon A_1(x) \ ,
\eea
and we change the basis of MI's via the Magnus series  obtained by using   $A_0$
as kernel,
\bea
f(\epsilon,x) = B_0(x) \  g(\epsilon,x) \ , \qquad  B_0(x) \equiv e^{ \magnus{A_0} (x,x_0) } \,  .
\label{Eq:change0}
\label{eq:MagnusA0}
\eea
Using Eq.~(\ref{eq:dexpOmega}), one can show that $B_0$ obeys the equation,
\bea
\partial_x B_0(x) = A_0(x) B_0(x) \ ,
\label{eq:dotB}
\eea 
which, analogously to the quantum-mechanical case, Eq.~(\ref{Eq:eqB}),
implies that the new basis $g$ of MI's fulfills a
system of differential equations  in the {\it canonical} factorized form,
\bea
\partial_x g(\epsilon,x) = \epsilon {\hat A}_1(x) g(\epsilon,x) \,  .
\label{eq:cansys}
\eea
The matrix ${\hat A}_1$ is related to  $A_1$ by a similarity map,
\bea
{\hat A}_1(x) = B_0^{-1}(x) A_1(x) B_0(x) \ ,
\label{eq:cantfm}
\eea
and does not depend on $\eps$.
The solution of Eq.~(\ref{eq:cansys}) can be found by using the Magnus theorem 
with $\epsilon \hat A_1$ as kernel
\bea
g(\epsilon,x) = B_1(\eps,x)  g_0(\epsilon) \ ,  \qquad  B_1(\eps,x)=e^{\magnus{\epsilon \hat A_1}(x,x_0)}  \, , 
\label{eq:MagnusA1} 
\eea
where the vector $g_0$ corresponds to the boundary values of the MI's.  
Therefore, the solution of the original system Eq.~(\ref{eq:MIdiffeq})
finally reads,
\bea
f (\epsilon,x)   =B_0(x)   B_1(\eps,x)  g_0(\epsilon) \, .
\eea

\noindent
It is worth to notice 
that $\magnus{\epsilon \hat A_1}$ in Eq.~(\ref{eq:MagnusA1}) depends on
$\epsilon$, while $\magnus{A_0}$ in Eq.~(\ref{eq:MagnusA0}) 
does not.  

Let us remark that the previously described two-step procedure 
is equivalent to solving, first, the {\it homogeneous} system 
 \begin{equation}
 \partial_x f_{\mbox{\tiny H}}(\epsilon,x) = A_0(x) f_{\mbox{\tiny H}}(\epsilon,x) \, , 
\end{equation}
whose solution reads,
\begin{equation}
  f_{\mbox{\tiny H}}(\epsilon,x) = B_0(x) g(\epsilon) \ ,
 \end{equation} 
 and,  then, to find the solution of the full system by Euler
 constants' variation. In fact, by promoting $g$ to be function of $x$,
 \begin{equation}
f_{\mbox{\tiny H}}(\epsilon,x) \rightarrow  f(\epsilon,x) =  B_0(x) g(\epsilon,x)  \ ,
\end{equation}
and by requiring $f$ to be solution of Eq.~(\ref{eq:MIdiffeq}), one
finds that $g(\epsilon,x)$ obeys the differential equation~(\ref{eq:cansys}).
\medskip

\noindent
The matrix $B_0$,  implementing the transformation from the
linear to the canonical form, is simply given as the
product of two matrix exponentials. Indeed one can split $A_0$
into a diagonal  term, $D_0$, and a matrix with vanishing diagonal entries $N_0$,
\bea
A_0(x) = D_0(x) + N_0(x) \ .
\eea
The transformation $B$ is then  obtained by the composition of two   transformations
\bea
B(x) =  e^{\magnus{D_0}(x,x_0)} e^{\magnus{\hat N_0}(x,x_0)}    =e^{\int_{x_0}^x d \tau  \ D_0(\tau)} e^{\magnus{\hat N_0}(x,x_0)} \ , 
\label{Eq:Bversion2}
\eea
where $\hat{N}_0$ is given by
\bea
\hat N_0(x) =    e^{-  \int_{x_0}^x d \tau \ D_0(\tau)}  \ N_0(x) \ e^{\int_{x_0}^x d \tau \ D_0(\tau)} 
\eea
In the last step of Eq.~(\ref{Eq:Bversion2}) we have used the  commutativity of the diagonal matrix $D_0$
with its own integral.  The leftmost expansion performs a transformation  that ``rotates" away $D_0$, while the
second expansion gets rid of the $\mathcal{O}(\epsilon^0)$ contribution coming from $\hat N_0$, i.e. coming from 
the image of $N_0$ under the first transformation.
 \medskip

\noindent
In the examples hereby discussed it is possible, by trials and errors, to find a set of MI's obeying a system of DE's linear in $\epsilon$.
Moreover in these cases  one finds that $\magnus{\hat N_0}$ 
contains just the first term of the series, except for the non-planar box,
where also the second order is non vanishing.

%% file: bhabha1L.tex
\section{One-Loop Bhabha scattering}
The calculation of the one-loop Bhabha scattering within  the DE's
method was discussed in \cite{BoboThesis,Bonciani:2003cj}, and more recently in
Ref.~\cite{Henn:2013woa}.  A selection of the Feynman diagrams contributing to this process
is depicted in Fig.~\ref{Fig:Diagrams1}.  In this section,  we  compute
a set of MI's with a slightly different definition from the ones in~\cite{Henn:2013woa},
which will be also adopted for the the one-loop $\times$ one-loop
subtopologies 
of the QED vertices in the next section.
\FIGURE[t]{
\input{PICTURES/Bhabha/Diagrams}

\caption{Selection of Feynman diagrams entering the Bhabha scattering at one loop.}
\label{Fig:Diagrams1}
}

The diagrams  depend on the invariants
$s=(p_1+p_2)^2$, $t=(p_1+p_3)^2$,  $u=(p_2+p_3)^2$  and on the fermion
mass $m$.  Momentum  conservation and the on-shellness 
of the external legs render these variables not independent as  they are related by the
condition $s+t+u = 4m^2$.   The integrals can be expressed in terms  of   the Landau
auxiliary variables $x$ and $y$,    defined as follows
\begin{equation}
s= -\frac{m^2 (1-x)^2}{x} \ , \qquad
t=-\frac{m^2 (1-y)^2}{y} \ .
\end{equation}
We identify the following basis $f$ of scalar integrals,
\begin{align}
f_1 &= \epsilon \top{1}\ ,  &
f_2 &= \epsilon \top{2}(t) \ ,   &
f_3 &= \epsilon \top{3}(s)  \ ,  \nonumber \\
f_4 &= \epsilon^2 \top{4}(t) \ ,  &
f_5 &= \epsilon^2 \top{5}(s,t)  \ , 
\end{align}
in terms of the integrals $\mathcal{T}$  in Fig.~\ref{Fig:Masters1}.
The basis $f$  fulfills the following systems of differential equations $(\sigma=x,y)$
\bea
\partial_\sigma f(\eps,x,y) = A_\sigma(\eps, x,y) f(\eps,x,y) \ , \quad A_\sigma(\eps,x,y) = D_{\sigma,0}(x,y) + \eps A_{\sigma,1}(x,y)  \ .
\eea
Both systems are linear in $\eps$ and in both cases the $\mathcal{O}(\eps^0)$ term, $D_{\sigma,0}$, 
is diagonal.  The systems can be brought in the canonical form by performing the
transformation 
\begin{equation}
f(\eps,x,y) = B_0(x,y) g(\eps,x,y) \qquad 
B_0(x,y)= e^{\int_{x_0}^x d \tau D_{ x,0}(\tau,y)}
 e^{\int_{y_0}^y d \tau D_{y,0}(x, \tau)} \ .
\end{equation}
The new basis $g$,
\begin{align}
g_1 &=  f_1  \ ,  &
g_2 &=  t \,f_2 \ , &
g_3 &=  \sqrt{\left(-s \right) \left(4 m^2-s\right)} \, f_3  \nonumber \\
g_4 &=  \sqrt{\left(-t \right) \left(4 m^2-t\right)} \, f_4  \ ,  &
g_5 &=  \sqrt{\left(-s \right) \left(4 m^2-s\right)} \, t \, f_5 \ . &
\end{align}
fulfills  the canonical systems
\begin{equation}
\partial_x g(\eps,x,y)=\epsilon \hat{A}_{x,1}(x,y)  \ g(\eps,x,y) \ , \qquad 
\partial_y g(\eps, x,y)=\epsilon \hat{A}_{y,1}(x,y) \ g(\eps,x,y) \ , 
\label{eq:bhabha:cansys}
\end{equation}
with
\begin{align}
\hat{A}_{x,1}(x,y) &= \left(
\begin{array}{ccccc}
 0 & 0 & 0 & 0 & 0 \\
 0 & 0 & 0 & 0 & 0 \\
 \frac{1}{x} & 0 & \frac{1-x}{x (1+x)} & 0 & 0 \\
 0 & 0 & 0 & 0 & 0 \\
 0 & \frac{2}{x} & \frac{2 (1-x) (1-y)^2}{(1+x) (x+y) (1+x y)} &
 -\frac{2 (1-y) (1+y)}{(x+y) (1+x y)} & 
 \frac{(1-x) (1-y)^2}{(1+x) (x+y) (1+x y)} \\
\end{array}
\right)  \, ,  \nonumber \\[0.5ex] \allowdisplaybreaks[1]
 \hat{A}_{y,1}(x,y) &=\left(
\begin{array}{ccccc}
 0 & 0 & 0 & 0 & 0 \\
 0 & \frac{1+y}{(1-y) y} & 0 & 0 & 0 \\
 0 & 0 & 0 & 0 & 0 \\ 
 \frac{1}{y} & \frac{1}{y} & 0 & \frac{4}{(1-y) (y+1)} & 0 \\
 0 & 0 & -\frac{2 x (1-y) (1+y)}{y (x+y) (1+x y)} & 
-\frac{2 (1-x) (1+x)}{(x+y) (1+x y)} & \frac{(1+x)^2 (1+y)}{(1-y) (x+y) (1+x y)} \\
\end{array}
\right)  \, . \allowdisplaybreaks[1]
\end{align}
\FIGURE[t]{
\input{PICTURES/Bhabha/Masters}

\caption{MI's for the one-loop corrections to the Bhabha scattering. All the external momenta are incoming. A dot denotes a squared propagator.}
\label{Fig:Masters1}
}
The two systems of DE's in Eq.(\ref{eq:bhabha:cansys})
can be combined in a full differential form, along the  lines of Ref.~\cite{Henn:2013woa}, 
\bea
d g(\eps, x,y)  = \eps \ d {\hat {\cal A}}_1(x,y) \ g(\eps, x,y) \ , 
\label{Eq:SystemBhabha}
\eea
where the matrix ${\hat{\cal A}}_1$ fulfills the relations,
\bea
 {\partial_x {\hat {\cal A}}_1(x,y)} = {\hat A}_{x,1}(x,y) \ , \qquad
  {\partial_y {\hat {\cal A}}_1(x,y)} = {\hat A}_{y,1}(x,y) \ .
 \eea
 and the integrability condition
 \bea
\eps \left (  \partial_x  \partial_y \hat {\cal A}_1(x,y)  -  \partial_y  \partial_x \hat {\cal A}_1(x,y) \right ) + \eps^2 \,  
 \left [  \partial_x \hat {\cal A}_1(x,y) ,  \partial_y \hat {\cal A}_1(x,y)  \right ] =0\ .
 \eea
The matrix  $\hat{{\cal A}}_1$ is  logarithmic in the variables $x$ and  $y$,
\bea
\hat{{\cal A}}_1(x,y)&=&
   M_1 \, \text{log}(x)
+M_2 \, \text{log}(1+x)
+ M_3 \, \text{log}(y) 
+ M_4 \, \text{log}(1+y) + \nn
&&
+ M_5 \, \text{log}(1-y)
+ M_6 \, \text{log}(x+y)
+ M_7 \, \text{log}(1+xy) \ ,
\eea
with 
\begin{align}
M_1 &=\left(
\begin{array}{ccccc}
 0 & 0 & 0 & 0 & 0 \\
 0 & 0 & 0 & 0 & 0 \\
 1 & 0 & 1 & 0 & 0 \\
 0 & 0 & 0 & 0 & 0 \\
 0 & 2 & 0 & 0 & 0 \\
\end{array}
\right) \ , &
M_2 &=\left(
\begin{array}{ccccc}
 0 & 0 & 0 & 0 & 0 \\
 0 & 0 & 0 & 0 & 0 \\
 0 & 0 & -2 & 0 & 0 \\
 0 & 0 & 0 & 0 & 0 \\
 0 & 0 & -4 & 0 & -2 \\
\end{array}
\right) \ ,  &
M_3 &=\left(
\begin{array}{ccccc}
 0 & 0 & 0 & 0 & 0 \\
 0 & 1 & 0 & 0 & 0 \\
 0 & 0 & 0 & 0 & 0 \\
 1 & 1 & 0 & 0 & 0 \\
 0 & 0 & -2 & 0 & 0 \\
\end{array}
\right) \ ,\nonumber \\[0.5ex] \allowdisplaybreaks[1]
M_4&=\left(
\begin{array}{ccccc}
 0 & 0 & 0 & 0 & 0 \\
 0 & 0 & 0 & 0 & 0 \\
 0 & 0 & 0 & 0 & 0 \\
 0 & 0 & 0 & 2 & 0 \\
 0 & 0 & 0 & 0 & 0 \\
\end{array}
\right)\ , &
M_5&=\left(
\begin{array}{ccccc}
 0 & 0 & 0 & 0 & 0 \\
 0 & -2 & 0 & 0 & 0 \\
 0 & 0 & 0 & 0 & 0 \\
 0 & 0 & 0 & -2 & 0 \\
 0 & 0 & 0 & 0 & 2 \\
\end{array}
\right)\ , &
M_6&=\left(
\begin{array}{ccccc}
 0 & 0 & 0 & 0 & 0 \\
 0 & 0 & 0 & 0 & 0 \\
 0 & 0 & 0 & 0 & 0 \\
 0 & 0 & 0 & 0 & 0 \\
 0 & 0 & -2 & -2 & -1 \\
\end{array}
\right)\ , \nonumber \\[0.5ex] \allowdisplaybreaks[1]
M_7&=\left(
\begin{array}{ccccc}
 0 & 0 & 0 & 0 & 0 \\
 0 & 0 & 0 & 0 & 0 \\
 0 & 0 & 0 & 0 & 0 \\
 0 & 0 & 0 & 0 & 0 \\
 0 & 0 & 2 & 2 & 1 \\
\end{array}
\right) \ .
\end{align}
The position of the non-zero  entries of the sparse matrices $M_i$ agrees with the result obtained in Ref.~\cite{Henn:2013woa}.
The actual value of the non-zero entries, however, are different, owing to the different normalization of the elements 
of the basis of MI's. The solution of the system~(\ref{Eq:SystemBhabha}) can be computed along the lines 
of Ref.~\cite{Henn:2013woa}. In particular, the solution is computed in the Euclidean region $0<x, y <1$
by using the analytic structures of the $g_i$ and then extended 
in the physical region by analytic continuation~\cite{Bonciani:2003cj}.

%% file: PICTURES/Bhabha/Diagrams.tex
\unitlength=0.25bp%
\begin{feynartspicture}(300,300)(1,1)
\FADiagram{}
\FAProp(0.,15.)(4.,10.)(0.,){Straight}{1}
\FAProp(0.,5.)(4.,10.)(0.,){Straight}{-1}
\FAProp(20.,15.)(16.,13.5)(0.,){Straight}{-1}
\FAProp(20.,5.)(16.,6.5)(0.,){Straight}{1}
\FAProp(4.,10.)(10.,10.)(0.,){Sine}{0}
\FAProp(16.,13.5)(16.,6.5)(0.,){Sine}{0}
\FAProp(16.,13.5)(10.,10.)(0.,){Straight}{-1}
\FAProp(16.,6.5)(10.,10.)(0.,){Straight}{1}
\end{feynartspicture}
\hspace{0.3cm}
%
%
\begin{feynartspicture}(300,300)(1,1)
\FADiagram{}
\FAProp(0.,15.)(6.5,13.5)(0.,){Straight}{1}
\FAProp(0.,5.)(6.5,6.5)(0.,){Straight}{-1}
\FAProp(20.,15.)(13.5,13.5)(0.,){Straight}{-1}
\FAProp(20.,5.)(13.5,6.5)(0.,){Straight}{1}
\FAProp(6.5,13.5)(6.5,6.5)(0.,){Sine}{0}
\FAProp(6.5,13.5)(13.5,13.5)(0.,){Straight}{1}
\FAProp(6.5,6.5)(13.5,6.5)(0.,){Straight}{-1}
\FAProp(13.5,13.5)(13.5,6.5)(0.,){Sine}{0}
\end{feynartspicture}
\hspace{0.3cm}
\begin{feynartspicture}(300,300)(1,1)
\FADiagram{}
\FAProp(0.,15.)(3.,10.)(0.,){Straight}{1}
\FAProp(0.,5.)(3.,10.)(0.,){Straight}{-1}
\FAProp(20.,15.)(17.,10.)(0.,){Straight}{-1}
\FAProp(20.,5.)(17.,10.)(0.,){Straight}{1}
\FAProp(3.,10.)(7.,10.)(0.,){Sine}{0}
\FAProp(17.,10.)(13.,10.)(0.,){Sine}{0}
\FAProp(7.,10.)(13.,10.)(0.8,){Straight}{-1}
\FAProp(7.,10.)(13.,10.)(-0.8,){Straight}{1}
\end{feynartspicture}

%% file: PICTURES/Bhabha/Masters.tex
\unitlength=0.25bp%
%
%
\begin{feynartspicture}(300,300)(1,1)
\FADiagram{}
\FAProp(10.,10.)(10., 15.)(.6,){Straight}{0}
\FAProp(10.,10.)(10., 15.)(-.6,){Straight}{0}
\FAVert(10.,15.){0}
\FALabel(10.,-1.)[]{$\top{1}$}
\end{feynartspicture}
\begin{feynartspicture}(300,300)(1,1)
\FADiagram{}
\FAProp(0.,15.)(10.,14.)(0.,){Straight}{0}
\FAProp(0.,5.)(10.,6.)(0.,){Straight}{0}
\FAProp(20.,15.)(10.,14.)(0.,){Straight}{0}
\FAProp(20.,5.)(10.,6.)(0.,){Straight}{0}
\FAProp(10.,14.)(10.,6.)(0.8,){Sine}{0}
\FAVert(13.2,10.){0}
\FAProp(10.,14.)(10.,6.)(-0.8,){Sine}{0}
\FALabel(5.2,16.3)[]{\small $p_1$}
\FALabel(5.2,3.7)[]{\small $p_2$}
\FALabel(14.8,16.3)[]{\small $p_3$}
\FALabel(14.8,3.7)[]{\small $p_4$}
\FALabel(10.,-1.)[]{$\top{2}(t)$}
\end{feynartspicture}
\begin{feynartspicture}(300,300)(1,1)
\FADiagram{}
\FAProp(0.,15.)(6.,10.)(0.,){Straight}{0}
\FAProp(0.,5.)(6.,10.)(0.,){Straight}{0}
\FAProp(20.,15.)(14.,10.)(0.,){Straight}{0}
\FAProp(20.,5.)(14.,10.)(0.,){Straight}{0}
\FAProp(6.,10.)(14.,10.)(0.8,){Straight}{0}
\FAVert(10.,13.2){0}
\FAProp(6.,10.)(14.,10.)(-0.8,){Straight}{0}
\FALabel(4.3,16.3)[]{\small $p_1$}
\FALabel(4.3,3.7)[]{\small $p_2$}
\FALabel(15.7,16.3)[]{\small $p_3$}
\FALabel(15.7,3.7)[]{\small $p_4$}
\FALabel(10.,-1.)[]{$\top{3}(s)$}
\end{feynartspicture}
\begin{feynartspicture}(300,300)(1,1)
\FADiagram{}
\FAProp(0.,15.)(6.5,13.)(0.,){Straight}{0}
\FAProp(0.,5.)(10.,7.)(0.,){Straight}{0}
\FAProp(20.,15.)(13.5,13.)(0.,){Straight}{0}
\FAProp(20.,5.)(10.,7.)(0.,){Straight}{0}
\FAProp(6.5,13.)(13.5,13.)(0.,){Straight}{0}
\FAProp(6.5,13.)(10.,7.)(0.,){Sine}{0}
\FAProp(13.5,13.)(10.,7.)(0.,){Sine}{0}
\FALabel(5.2,16.3)[]{\small $p_1$}
\FALabel(5.2,3.7)[]{\small $p_2$}
\FALabel(14.8,16.3)[]{\small $p_3$}
\FALabel(14.8,3.7)[]{\small $p_4$}
\FALabel(10.,-1.)[]{$\top{4}(t)$}
\end{feynartspicture}
\begin{feynartspicture}(300,300)(1,1)
\FADiagram{}
\FAProp(0.,15.)(6.5,13.5)(0.,){Straight}{0}
\FAProp(0.,5.)(6.5,6.5)(0.,){Straight}{0}
\FAProp(20.,15.)(13.5,13.5)(0.,){Straight}{0}
\FAProp(20.,5.)(13.5,6.5)(0.,){Straight}{0}
\FAProp(6.5,13.5)(6.5,6.5)(0.,){Sine}{0}
\FAProp(6.5,13.5)(13.5,13.5)(0.,){Straight}{0}
\FAProp(6.5,6.5)(13.5,6.5)(0.,){Straight}{0}
\FAProp(13.5,13.5)(13.5,6.5)(0.,){Sine}{0}
\FALabel(5.2,16.3)[]{\small $p_1$}
\FALabel(5.2,3.7)[]{\small $p_2$}
\FALabel(14.8,16.3)[]{\small $p_3$}
\FALabel(14.8,3.7)[]{\small $p_4$}
\FALabel(10.,-1.)[]{$\top{5}(s,t)$}
\end{feynartspicture}

%% file: vertex2L.tex
\FIGURE[t]{
\input{PICTURES/Vertex2/Diagrams}
\caption{Selection of Feynman diagrams entering the correction of the QED vertex at two loops. The  internal momenta in the
first diagram are oriented according to the fermion flow, while the external momenta are incoming.}
\label{Fig:Diagrams2}
}

\section{Two-Loop QED Vertices}
\label{Sec:vertex}
A basis of MI's for the electron form factor at two loops in QED \cite{Bonciani:2003ai}
was computed in Ref.~\cite{Bonciani:2003te}, for arbitrary kinematics and finite electron mass.
The diagrams contributing to such corrections are depicted in Fig.~\ref{Fig:Diagrams2} and
depend on $s=(p_1+p_2)^2$ and $p_1^2=p_2^2=m^2$. In this example we
start from an alternative set of MI's,
\begin{align}
f_1 &= \epsilon^2 \top{1} \ , &
f_2 &= \epsilon^2   \top{2}  \ , &
f_3 &= \epsilon^2  \top{3}  \ , & 
f_4 &= \epsilon^2 \top{4}  \ , &
f_5 &= \epsilon^2   \top{5}  \ , \nn \displaybreak[1]
f_6 &= \epsilon^2  \top{6}   \ , &
f_7 &= \epsilon^2  \top{7}  \ , &
f_8 &= \epsilon^3  \top{8}  \ , & 
f_9 &= \epsilon^3   \top{9}  \ , &
f_{10} &= \epsilon^2  \top{10}  \ ,  \nn\displaybreak[1]
f_{11} &= \epsilon^3  \top{11}  \ ,  &
f_{12} &= \epsilon^3  \top{12}  \ , &
f_{13} &= \epsilon^2    \top{13}  \ ,  &
f_{14} &= \epsilon^3  \top{14}   \ ,  &
f_{15} &= \epsilon^4   \top{15}  \ , \nn \displaybreak[1]
f_{16} &= \epsilon^4  \top{16}   \ ,  &
f_{17} &=  \epsilon^4  \top{17}  \ ,  &
\end{align}
where the integrals $\top{i}$ are collected in Fig.~\ref{Fig:Masters2}.
The system of differential equation for $f$, in the auxiliary
variable $x$, defined through 
 \begin{equation}
s= -\frac{m^2 (1-x)^2}{x} \, , 
\end{equation}
is linear in $\eps$,
\bea
\partial_x f(\eps,x) = A(\eps, x) \  f(\eps,x) \ , \qquad A(\eps, x) = A_{0}(x) + \eps A_{1}(x)  \ .
\eea
The canonical form can be obtained performing the transformation described in Section~\ref{Sec:diffeq},
\bea
f(\eps, x)  = B_0(x) \ g(\eps,x), \qquad B_0(x) = e^{\magnus{A_0}(x)  }  \ .
\eea
The new basis $g$ is given by
\begin{align}
g_1 &= f_1 \ , &   g_2 &=\lambda_1 f_2\ ,  \nn
 g_3 &=  (-s)\lambda_2 f_3 \ , & g_4 &=  m^2 f_4\ , \nn
 g_{5} &=\lambda_1  \left ( f_5 +\frac{f_6}{2} \right ) - \frac{s}{2} f_6  \ ,& g_6 &=   (-s) f_6 \ ,\nn
g_7 &=  m^2 f_7 \ ,&   g_8 &=  \lambda_1 f_8 \ ,\nn
g_9 &=   \lambda_1 f_9  \ ,& 
g_{10} &= \lambda_3  \left ( 2  f_5 +f_6\right ) +  m^2 \lambda_2 f_{10} \ , \nn
g_{11} &=  \lambda_1 f_{11} \ , &  g_{12} &=  \lambda_1 f_{12} \ ,\nn
g_{13} &=  3\, \left( m^2-\frac{ s}{2}\right) f_7 -  s \lambda_2 f_{13} \ , & g_{14} &=   (-s) \lambda_2 f_{14}\ , \nn
 g_{15}&=  \lambda_1 f_{15} \ ,& g_{16} &=  \lambda_1 f_{16} \ ,\nn
  g_{17} &= (-s) \lambda_2 f_{17} \ ,
  \label{Eq:MIfinVertex}
\end{align}
where 
\bea
\lambda_1 =   \sqrt{-s}\sqrt{4m^2-s} \ ,   \qquad \lambda_2 =  (4m^2-s) \ ,   \qquad \lambda_3  = \frac{\lambda_1+\lambda_2}{4} \ .
\eea
The new basis of MI's obeys a system of DE's in the canonical form,
\FIGURE[t]{
\input{PICTURES/Vertex2/Masters}
\caption{MI's for the two-loop corrections to the QED vertex. All the external momenta depicted are incoming. In the integral  $\top{16}$ 
the loop momenta $k_1$, $k_2$ are fixed according to the first diagram of Fig.~\ref{Fig:Diagrams2} and a term
$(k_1+k_2)^2$ has to be included in  the numerator of the integrand. A dot indicates a squared propagator.}
\label{Fig:Masters2}
}
\begin{equation}
\partial_x g(\eps, x) =\epsilon \hat{A}_1(x) g(\eps, x) \ , \qquad \hat{A}_1(x)= \frac{M_1}{x}+ \frac{M_2}{1+x} + \frac{M_3}{1-x} \, ,
\label{Eq:SystemVertex}
\end{equation}
with
\begin{align}
M_1 &= \scriptsize \left(
\begin{array}{ccccccccccccccccc}
 0 & 0 & 0 & 0 & 0 & 0 & 0 & 0 & 0 & 0 & 0 & 0 & 0 & 0 & 0 & 0 & 0
   \\
 1 & 1 & 0 & 0 & 0 & 0 & 0 & 0 & 0 & 0 & 0 & 0 & 0 & 0 & 0 & 0 & 0
   \\
 0 & 2 & 2 & 0 & 0 & 0 & 0 & 0 & 0 & 0 & 0 & 0 & 0 & 0 & 0 & 0 & 0
   \\
 0 & 0 & 0 & 0 & 0 & 0 & 0 & 0 & 0 & 0 & 0 & 0 & 0 & 0 & 0 & 0 & 0
   \\
 -1 & 0 & 0 & 0 & 5 & -6 & 0 & 0 & 0 & 0 & 0 & 0 & 0 & 0 & 0 & 0 &
   0 \\
 0 & 0 & 0 & 0 & 2 & -2 & 0 & 0 & 0 & 0 & 0 & 0 & 0 & 0 & 0 & 0 & 0
   \\
 0 & 0 & 0 & 0 & 0 & 0 & 0 & 0 & 0 & 0 & 0 & 0 & 0 & 0 & 0 & 0 & 0
   \\
 -1 & 0 & 0 & -4 & 0 & -2 & 0 & -2 & 0 & 0 & 0 & 0 & 0 & 0 & 0 & 0
   & 0 \\
 0 & 0 & 0 & -2 & 0 & 0 & 0 & 0 & 2 & 0 & 0 & 0 & 0 & 0 & 0 & 0 & 0
   \\
 -\frac{1}{2} & 0 & 0 & 0 & 1 & -2 & -3 & 0 & 0 & 3 & 3 & 0 & 0 & 0
   & 0 & 0 & 0 \\
 0 & 0 & 0 & 0 & 1 & -1 & 2 & 0 & 0 & -2 & -2 & 0 & 0 & 0 & 0 & 0 &
   0 \\
 0 & 0 & 0 & 0 & 0 & 0 & 1 & 0 & 0 & 0 & 0 & -1 & -1 & 0 & 0 & 0 &
   0 \\
 0 & -1 & 0 & 0 & 0 & 0 & -3 & 0 & 0 & 0 & 0 & 3 & 3 & 0 & 0 & 0 &
   0 \\
 0 & -1 & 0 & 0 & 1 & -\frac{1}{2} & 0 & 2 & 2 & 0 & 0 & 0 & 0 & 2
   & 2 & 0 & 0 \\
 0 & 0 & 0 & 0 & 0 & \frac{1}{2} & 0 & -\frac{1}{2} & 0 & 0 & 0 & 0
   & 0 & -1 & -1 & 0 & 0 \\
 -\frac{1}{2} & 0 & 0 & -2 & -1 & 0 & -2 & 1 & 0 & 2 & 0 & -2 & 0 &
   0 & -2 & -2 & 2 \\
 0 & 0 & 0 & 0 & -1 & \frac{1}{2} & 0 & 3 & -2 & 0 & -6 & -2 & 0 &
   0 & -4 & -4 & 4 \\
\end{array}
\right)  \normalsize \ , \nn[0.5ex] \displaybreak[1]
M_2 &=  \scriptsize \left(
\begin{array}{ccccccccccccccccc}
 0 & 0 & 0 & 0 & 0 & 0 & 0 & 0 & 0 & 0 & 0 & 0 & 0 & 0 & 0 & 0 & 0
   \\
 0 & -2 & 0 & 0 & 0 & 0 & 0 & 0 & 0 & 0 & 0 & 0 & 0 & 0 & 0 & 0 & 0
   \\
 0 & 0 & -4 & 0 & 0 & 0 & 0 & 0 & 0 & 0 & 0 & 0 & 0 & 0 & 0 & 0 & 0
   \\
 0 & 0 & 0 & 0 & 0 & 0 & 0 & 0 & 0 & 0 & 0 & 0 & 0 & 0 & 0 & 0 & 0
   \\
 0 & 0 & 0 & 0 & -6 & 3 & 0 & 0 & 0 & 0 & 0 & 0 & 0 & 0 & 0 & 0 & 0
   \\
 0 & 0 & 0 & 0 & 0 & 0 & 0 & 0 & 0 & 0 & 0 & 0 & 0 & 0 & 0 & 0 & 0
   \\
 0 & 0 & 0 & 0 & 0 & 0 & 0 & 0 & 0 & 0 & 0 & 0 & 0 & 0 & 0 & 0 & 0
   \\
 0 & 0 & 0 & 0 & 0 & 0 & 0 & 2 & 0 & 0 & 0 & 0 & 0 & 0 & 0 & 0 & 0
   \\
 0 & 0 & 0 & 0 & 0 & 0 & 0 & 0 & -4 & 0 & 0 & 0 & 0 & 0 & 0 & 0 & 0
   \\
 0 & 0 & 0 & 0 & -1 & \frac{1}{2} & 0 & 0 & 0 & -4 & 0 & 0 & 0 & 0
   & 0 & 0 & 0 \\
 0 & 0 & 0 & 0 & 0 & 0 & 0 & 0 & 0 & 0 & 2 & 0 & 0 & 0 & 0 & 0 & 0
   \\
 0 & 0 & 0 & 0 & 0 & 0 & 0 & 0 & 0 & 0 & 0 & 2 & 0 & 0 & 0 & 0 & 0
   \\
 0 & 0 & 0 & 0 & 0 & 0 & -6 & 0 & 0 & 0 & 0 & 0 & -2 & 0 & 0 & 0 &
   0 \\
 0 & 0 & 0 & 0 & 0 & 0 & 0 & 0 & 0 & 0 & 0 & 0 & 0 & -4 & 0 & 0 & 0
   \\
 0 & 0 & 0 & 0 & 0 & 0 & 0 & 0 & 0 & 0 & 0 & 0 & 0 & 0 & 2 & 0 & 0
   \\
 0 & 0 & 0 & 0 & 0 & 0 & 0 & 0 & 0 & 0 & 0 & 0 & 0 & 0 & 0 & 2 & 0
   \\
 0 & 0 & 0 & 0 & 0 & 0 & 0 & 0 & 0 & 0 & 0 & 0 & 0 & 0 & 0 & 0 & -4
   \\
\end{array}
\right) \ ,\normalsize  \nn  [0.5ex] \displaybreak[1]
M_3&=  \scriptsize \left(
\begin{array}{ccccccccccccccccc}
 0 & 0 & 0 & 0 & 0 & 0 & 0 & 0 & 0 & 0 & 0 & 0 & 0 & 0 & 0 & 0 & 0
   \\
 0 & 0 & 0 & 0 & 0 & 0 & 0 & 0 & 0 & 0 & 0 & 0 & 0 & 0 & 0 & 0 & 0
   \\
 0 & 0 & 0 & 0 & 0 & 0 & 0 & 0 & 0 & 0 & 0 & 0 & 0 & 0 & 0 & 0 & 0
   \\
 0 & 0 & 0 & 0 & 0 & 0 & 0 & 0 & 0 & 0 & 0 & 0 & 0 & 0 & 0 & 0 & 0
   \\
 0 & 0 & 0 & 0 & 2 & -2 & 0 & 0 & 0 & 0 & 0 & 0 & 0 & 0 & 0 & 0 & 0
   \\
 0 & 0 & 0 & 0 & 0 & -2 & 0 & 0 & 0 & 0 & 0 & 0 & 0 & 0 & 0 & 0 & 0
   \\
 0 & 0 & 0 & 0 & 0 & 0 & 0 & 0 & 0 & 0 & 0 & 0 & 0 & 0 & 0 & 0 & 0
   \\
 0 & 0 & 0 & 0 & 0 & 0 & 0 & -2 & 0 & 0 & 0 & 0 & 0 & 0 & 0 & 0 & 0
   \\
 0 & 0 & 0 & 0 & 0 & 0 & 0 & 0 & 0 & 0 & 0 & 0 & 0 & 0 & 0 & 0 & 0
   \\
 0 & 0 & 0 & 0 & 0 & 0 & -6 & 0 & 0 & 2 & 0 & 0 & 0 & 0 & 0 & 0 & 0
   \\
 0 & 0 & 0 & 0 & 0 & 0 & 0 & 0 & 0 & 0 & -2 & 0 & 0 & 0 & 0 & 0 & 0
   \\
 0 & 0 & 0 & 0 & 0 & 0 & 0 & 0 & 0 & 0 & 0 & 0 & 0 & 0 & 0 & 0 & 0
   \\
 0 & 0 & 0 & 0 & 0 & 0 & -12 & 0 & 0 & 0 & 0 & 0 & 4 & 0 & 0 & 0 &
   0 \\
 0 & 0 & 0 & 0 & 0 & 0 & 0 & 0 & 0 & 0 & 0 & 0 & 0 & 0 & 0 & 0 & 0
   \\
 0 & 0 & 0 & 0 & 0 & 0 & 0 & -1 & 0 & 0 & 0 & 0 & 0 & 0 & 0 & 0 & 0
   \\
 0 & 0 & 0 & 0 & 0 & 0 & 0 & 2 & 0 & 0 & 0 & -4 & 0 & 0 & -4 & -2 & 0
   \\
 0 & 0 & 0 & 0 & 0 & 0 & 0 & 0 & 0 & 0 & 0 & 0 & 0 & 0 & 0 & 0 & 4
   \\
\end{array}
\right) \ . \normalsize \displaybreak[1]
\end{align}
The solution of the system can be expressed as Dyson series, as well
as Magnus series, in terms
of one-dimensional  Harmonic Polylogarithms  (HPL's)~\cite{Remiddi:1999ew}. The requirements
that the MI's are real-valued in the Euclidean region and regular  
in $x=1$ ($s=0$), or simply the matching against the known integrals
at $x=1$, fix all but three boundary conditions, 
corresponding to the {\it constant} MI's 
$g_1$, $g_4$ and  $g_7$ (that do not depend on $x$).   The integrals $g_1$ and $g_4$ can be easily
computed by direct integration, while 
$g_7$ can be determined from the results of Ref.~\cite{Argeri:2002wz}. Our results were checked analytically, 
using the code {\sc HPL}~\cite{Maitre:2005uu, Maitre:2007kp},  against 
the results available in the literature~\cite{Bonciani:2003te}.
The expressions of the transcendentally homogenous MI's $g$ are shown
in Appendix~\ref{results:vertex2L}, and collected in the ancillary file {\tt <vertex2L.m>}.

%% file: PICTURES/Vertex2/Diagrams.tex
\unitlength=0.27bp%
\begin{feynartspicture}(300,300)(1,1)
\FADiagram{}
\FAProp(0.,15.)(5.,15.)(0.,){Straight}{1}
\FAProp(0.,5.)(5.,5.)(0.,){Straight}{-1}
\FAProp(20.,10.)(14.5,10.)(0.,){Sine}{0}
\FAProp(9.,12.95)(5.,5.)(0.,){Sine}{0}
\FAProp(9.,12.95)(5.,15.)(0.,){Straight}{-1}
\FAProp(9.,12.95)(14.5,10.)(0.,){Straight}{1}
\FAProp(9.,7.1)(5.,15)(0.,){Sine}{0}
\FAProp(9.,7.1)(5.,5.)(0.,){Straight}{1}
\FAProp(9.,7.1)(14.5,10.)(0.,){Straight}{-1}
%
%
\FALabel(2.5,16.7)[b]{\small $p_1$}
\FALabel(2.5,6.7)[b]{\small $p_2$}
\FALabel(11.5,14.8)[]{\scriptsize $p_1-k_1$}
\FALabel(11.5,5.2)[]{\scriptsize $-p_2-k_2$}
\end{feynartspicture}
\begin{feynartspicture}(300,300)(1,1)
\FADiagram{}
\FAProp(0.,15.)(5.,15.)(0.,){Straight}{1}
\FAProp(0.,5.)(5.,5.)(0.,){Straight}{-1}
\FAProp(20.,10.)(14.5,10.)(0.,){Sine}{0}
\FAProp(9.,12.95)(9.,7.1)(0.,){Sine}{0}
\FAProp(9.,12.95)(5.,15.)(0.,){Straight}{-1}
\FAProp(9.,12.95)(14.5,10.)(0.,){Straight}{1}
\FAProp(5.,5.)(5.,15)(0.,){Sine}{0}
\FAProp(9.,7.1)(5.,5.)(0.,){Straight}{1}
\FAProp(9.,7.1)(14.5,10.)(0.,){Straight}{-1}
\end{feynartspicture}
\begin{feynartspicture}(300,300)(1,1)
\FADiagram{}
\FAProp(0.,15.)(5.,15.)(0.,){Straight}{1}
\FAProp(0.,5.)(5.,5.)(0.,){Straight}{-1}
\FAProp(20.,10.)(14.5,10.)(0.,){Sine}{0}
\FAProp(10.45,12.1)(5.,15.)(0.,){Sine}{0}
\FAProp(10.45,12.1)(14.5,10.)(0.,){Straight}{1}
\FAProp(5.,9.)(10.45,12.1)(0.,){Straight}{1}
\FAProp(5.,9.)(5.,15.)(0.,){Straight}{-1}
\FAProp(5.,9.)(5.,5.)(0.,){Sine}{0}
\FAProp(5.,5.)(14.5,10.)(0.,){Straight}{-1}
\end{feynartspicture}
\begin{feynartspicture}(300,300)(1,1)
\FADiagram{}
\FAProp(0.,15.)(5.,15.)(0.,){Straight}{1}
\FAProp(0.,5.)(5.,5.)(0.,){Straight}{-1}
\FAProp(20.,10.)(14.5,10.)(0.,){Sine}{0}
\FAProp(5.,15.)(5.,5.)(0.,){Sine}{0}
\FAProp(5.,15.)(14.5,10.)(0.,){Straight}{1}
\FAProp(7.75,6.4)(5.,5.)(0.,){Straight}{1}
\FAProp(11.9,8.65)(7.75,6.4)(-0.8,){Straight}{1}
\FAProp(11.9,8.65)(7.75,6.4)(0.8,){Sine}{0}
\FAProp(11.9,8.65)(14.5,10.)(0.,){Straight}{-1}
\end{feynartspicture}
\begin{feynartspicture}(300,300)(1,1)
\FADiagram{}
\FAProp(0.,15.)(5.5,15.)(0.,){Straight}{1}
\FAProp(0.,5.)(5.5,5.)(0.,){Straight}{-1}
\FAProp(20.,10.)(14.5,10.)(0.,){Sine}{0}
\FAProp(5.5,15.)(14.5,10.)(0.,){Straight}{1}
\FAProp(5.5,12.5)(5.5,15.)(0.,){Sine}{0}
\FAProp(5.5,7.5)(5.5,12.5)(-0.8,){Straight}{-1}
\FAProp(5.5,7.5)(5.5,12.5)(0.8,){Straight}{1}
\FAProp(5.5,7.5)(5.5,5.)(0.,){Sine}{0}
\FAProp(5.5,5.)(14.5,10.)(0.,){Straight}{-1}
\end{feynartspicture}

%% file: PICTURES/Vertex2/Masters.tex
\unitlength=0.25bp%
%
\begin{feynartspicture}(300,300)(1,1)
\FADiagram{}
\FAProp(10.,10.)(10., 15.)(.6,){Straight}{0}
\FAProp(10.,10.)(10., 15.)(-.6,){Straight}{0}
\FAVert(10.,15.){0}
\FAProp(10.,10.)(10., 5.)(.6,){Straight}{0}
\FAProp(10.,10.)(10., 5.)(-.6,){Straight}{0}
\FAVert(10.,5.){0}
\FALabel(10.,0.)[]{$\top{1}$}
\end{feynartspicture}

%
\begin{feynartspicture}(300,300)(1,1)
\FADiagram{}
\FAProp(0.,10.)(6.,10.)(0.,){Straight}{0}
\FAProp(20.,10.)(14.,10.)(0.,){Straight}{0}
\FAProp(14.,10.)(6.,10.)(-1.,){Straight}{0}
\FAVert(10.,14.){0}
\FAProp(14.,10.)(6.,10.)(1.,){Straight}{0}
\FAProp(14.,10.)(17.5355, 13.5355)(.6,){Straight}{0}
\FAProp(14.,10.)(17.5355, 13.5355)(-.6,){Straight}{0}
\FAVert(17.5355, 13.5355){0}
\FALabel(2.,8.93)[t]{\small $p_{12}$}
\FALabel(10.,0.)[]{$\top{2}(s)$}
\end{feynartspicture}
%
\begin{feynartspicture}(300,300)(1,1)
\FADiagram{}
\FAProp(0.,10.)(4.,10.)(0.,){Straight}{0}
\FAProp(20.,10.)(16.,10.)(0.,){Straight}{0}
\FAProp(4.,10.)(10.,10.)(1.,){Straight}{0}
\FAProp(4.,10.)(10.,10.)(-1.,){Straight}{0}
\FAProp(16.,10.)(10.,10.)(1.,){Straight}{0}
\FAProp(16.,10.)(10.,10.)(-1.,){Straight}{0}
\FAVert(7.,13.){0}
\FAVert(13.,13.){0}
\FALabel(2.,8.93)[t]{\small $p_{12}$}
\FALabel(10.,0.)[]{$\top{3}(s)$}
\end{feynartspicture}
%
\begin{feynartspicture}(300,300)(1,1)
\FADiagram{}
\FAProp(0.,10.)(6.,10.)(0.,){Straight}{0}
\FAProp(20.,10.)(14.,10.)(0.,){Straight}{0}
\FAProp(14.,10.)(6.,10.)(-1.,){Sine}{0}
\FAProp(14.,10.)(6.,10.)(0.,){Straight}{0}
\FAVert(10.,10.){0}
\FAVert(10.,6.){0}
\FAProp(14.,10.)(6.,10.)(1.,){Sine}{0}
\FALabel(2.,8.93)[t]{\small $p_1$}
\FALabel(10.,0.)[]{$\top{4}$}
\end{feynartspicture}
%
\begin{feynartspicture}(300,300)(1,1)
\FADiagram{}
\FAProp(0.,10.)(6.,10.)(0.,){Straight}{0}
\FAProp(20.,10.)(14.,10.)(0.,){Straight}{0}
\FAProp(14.,10.)(6.,10.)(-1.,){Sine}{0}
\FAProp(14.,10.)(6.,10.)(0.,){Straight}{0}
\FAVert(10.,10.){0}
\FAVert(10.,6.){0}
\FAProp(14.,10.)(6.,10.)(1.,){Straight}{0}
\FALabel(2.,8.93)[t]{\small $p_{12}$}
\FALabel(10.,0.)[]{$\top{5}(s)$}
\end{feynartspicture}
%
\begin{feynartspicture}(300,300)(1,1)
\FADiagram{}
\FAProp(0.,10.)(6.,10.)(0.,){Straight}{0}
\FAProp(20.,10.)(14.,10.)(0.,){Straight}{0}
\FAProp(14.,10.)(6.,10.)(-1.,){Sine}{0}
\FAProp(14.,10.)(6.,10.)(0.,){Straight}{0}
\FAVert(10.,10.){0}
\FAVert(10.,14.){0}
\FAProp(14.,10.)(6.,10.)(1.,){Straight}{0}
\FALabel(2.,8.93)[t]{\small $p_{12}$}
\FALabel(10.,0.)[]{$\top{6}(s)$}
\end{feynartspicture}
%
%
\begin{feynartspicture}(300,300)(1,1)
\FADiagram{}
\FAProp(0.,10.)(6.,10.)(0.,){Straight}{0}
\FAProp(20.,10.)(14.,10.)(0.,){Straight}{0}
\FAProp(14.,10.)(6.,10.)(-1.,){Straight}{0}
\FAProp(14.,10.)(6.,10.)(0.,){Straight}{0}
\FAVert(10.,10.){0}
\FAVert(10.,6.){0}
\FAProp(14.,10.)(6.,10.)(1.,){Straight}{0}
\FALabel(2.,8.93)[t]{\small $p_{2}$}
\FALabel(10.,0.)[]{$\top{7}$}
\end{feynartspicture}
%
\begin{feynartspicture}(300,300)(1,1)
\FADiagram{}
\FAProp(0.,15.)(5.,15.)(0.,){Straight}{0}
\FAProp(0.,5.)(5.,5.)(0.,){Straight}{0}
\FAProp(20.,10.)(14.5,10.)(0.,){Straight}{0}
\FAProp(5.,5.)(14.5,10.)(0.,){Straight}{0}
\FAProp(5.,5.)(5.,15.)(0.,){Sine}{0}
\FAProp(14.5,10.)(5.,15.)(-0.410984,){Sine}{0}
\FAProp(14.5,10.)(5.,15.)(0.410984,){Straight}{0}
\FAVert(10.7775, 14.4522){0}
\FALabel(2.5,16.77)[b]{\small $p_1$}
\FALabel(2.5,6.77)[b]{\small $p_2$}
\FALabel(10.,0.)[]{$\top{8}(s)$}
\end{feynartspicture}
%
\begin{feynartspicture}(300,300)(1,1)
\FADiagram{}
\FAProp(0.,15.)(6.,15.)(0.,){Straight}{0}
\FAProp(0.,5.)(6.,5.)(0.,){Straight}{0}
\FAProp(20.,10.)(14.5,10.)(0.,){Straight}{0}
\FAProp(14.5,10.)(6.,5.)(0.,){Straight}{0}
\FAProp(14.5,10.)(6.,15.)(0.,){Straight}{0}
\FAProp(6.,5.)(6.,15.)(-0.4,){Sine}{0}
\FAProp(6.,5.)(6.,15.)(0.4,){Sine}{0}
\FAVert(8.,10.){0}
\FALabel(2.5,16.77)[b]{\small $p_1$}
\FALabel(2.5,6.77)[b]{\small $p_2$}
\FALabel(10.,0.)[]{$\top{9}(s)$}
\end{feynartspicture}
%
\begin{feynartspicture}(300,300)(1,1)
\FADiagram{}
\FAProp(0.,15.)(5.,15.)(0.,){Straight}{0}
\FAProp(0.,5.)(5.,5.)(0.,){Straight}{0}
\FAProp(20.,10.)(14.5,10.)(0.,){Straight}{0}
\FAProp(5.,5.)(14.5,10.)(0.,){Sine}{0}
\FAProp(5.,5.)(5.,15.)(0.,){Straight}{0}
\FAVert(5.,10.){0}
\FAProp(14.5,10.)(5.,15.)(-0.410984,){Straight}{0}
\FAProp(14.5,10.)(5.,15.)(0.410984,){Straight}{0}
\FAVert(10.7775, 14.4522){0}
\FALabel(2.5,16.77)[b]{\small $p_1$}
\FALabel(2.5,6.77)[b]{\small $p_2$}
\FALabel(10.,0.)[]{$\top{10}(s)$}
\end{feynartspicture}
%
\begin{feynartspicture}(300,300)(1,1)
\FADiagram{}
\FAProp(0.,15.)(5.,15.)(0.,){Straight}{0}
\FAProp(0.,5.)(5.,5.)(0.,){Straight}{0}
\FAProp(20.,10.)(14.5,10.)(0.,){Straight}{0}
\FAProp(5.,5.)(14.5,10.)(0.,){Sine}{0}
\FAProp(5.,5.)(5.,15.)(0.,){Straight}{0}
\FAProp(14.5,10.)(5.,15.)(-0.410984,){Straight}{0}
\FAProp(14.5,10.)(5.,15.)(0.410984,){Straight}{0}
\FAVert(10.7775, 14.4522){0}
\FALabel(2.5,16.77)[b]{\small $p_1$}
\FALabel(2.5,6.77)[b]{\small $p_2$}
\FALabel(10.,0.)[]{$\top{11}(s)$}
\end{feynartspicture}
%
\begin{feynartspicture}(300,300)(1,1)
\FADiagram{}
\FAProp(0.,15.)(6.,15.)(0.,){Straight}{0}
\FAProp(0.,5.)(6.,5.)(0.,){Straight}{0}
\FAProp(20.,10.)(14.5,10.)(0.,){Straight}{0}
\FAProp(14.5,10.)(6.,5.)(0.,){Straight}{0}
\FAProp(14.5,10.)(6.,15.)(0.,){Straight}{0}
\FAProp(6.,5.)(6.,15.)(-0.4,){Straight}{0}
\FAProp(6.,5.)(6.,15.)(0.4,){Straight}{0}
\FAVert(8.,10.){0}
\FALabel(2.5,16.77)[b]{\small $p_1$}
\FALabel(2.5,6.77)[b]{\small $p_2$}
\FALabel(10.,0.)[]{$\top{12}(s)$}
\end{feynartspicture}
%
\begin{feynartspicture}(300,300)(1,1)
\FADiagram{}
\FAProp(0.,15.)(6.,15.)(0.,){Straight}{0}
\FAProp(0.,5.)(6.,5.)(0.,){Straight}{0}
\FAProp(20.,10.)(14.5,10.)(0.,){Straight}{0}
\FAProp(14.5,10.)(6.,5.)(0.,){Straight}{0}
\FAProp(14.5,10.)(6.,15.)(0.,){Straight}{0}
\FAVert(10.25, 7.5){0}
\FAProp(6.,5.)(6.,15.)(-0.4,){Straight}{0}
\FAProp(6.,5.)(6.,15.)(0.4,){Straight}{0}
\FAVert(8.,10.){0}
\FALabel(2.5,16.77)[b]{\small $p_1$}
\FALabel(2.5,6.77)[b]{\small $p_2$}
\FALabel(10.,0.)[]{$\top{13}(s)$}
\end{feynartspicture}
%
\begin{feynartspicture}(300,300)(1,1)
\FADiagram{}
\FAProp(0.,15.)(5.5,15.)(0.,){Straight}{0}
\FAProp(0.,5.)(5.5,5.)(0.,){Straight}{0}
\FAProp(20.,10.)(15.,10.)(0.,){Straight}{0}
\FAProp(10.3,12.45)(5.5,15.)(0.,){Straight}{0}
\FAProp(10.3,12.45)(15.,10.)(0.,){Straight}{0}
\FAProp(10.3,12.45)(5.5,5.)(0.,){Sine}{0}
\FAProp(5.5,15.)(5.5,5.)(0.,){Sine}{0}
\FAProp(15.,10.)(5.5,5.)(0.,){Straight}{0}
\FAVert(10.25, 7.5){0}
\FALabel(2.5,16.77)[b]{\small $p_1$}
\FALabel(2.5,6.77)[b]{\small $p_2$}
\FALabel(10.,0.)[]{$\top{14}(s)$}
\end{feynartspicture}
%
\begin{feynartspicture}(300,300)(1,1)
\FADiagram{}
\FAProp(0.,15.)(5.5,15.)(0.,){Straight}{0}
\FAProp(0.,5.)(5.5,5.)(0.,){Straight}{0}
\FAProp(20.,10.)(15.,10.)(0.,){Straight}{0}
\FAProp(10.3,12.45)(5.5,15.)(0.,){Straight}{0}
\FAProp(10.3,12.45)(15.,10.)(0.,){Straight}{0}
\FAProp(10.3,12.45)(5.5,5.)(0.,){Sine}{0}
\FAProp(5.5,15.)(5.5,5.)(0.,){Sine}{0}
\FAProp(15.,10.)(5.5,5.)(0.,){Straight}{0}
\FALabel(2.5,16.77)[b]{\small $p_1$}
\FALabel(2.5,6.77)[b]{\small $p_2$}
\FALabel(10.,0.)[]{$\top{15}(s)$}
\end{feynartspicture}
%
%
\begin{feynartspicture}(300,300)(1,1)
\FADiagram{}
\FAProp(0.,15.)(5.,15.)(0.,){Straight}{0}
\FAProp(0.,5.)(5.,5.)(0.,){Straight}{0}
\FAProp(20.,10.)(14.5,10.)(0.,){Straight}{0}
\FAProp(5.,15.)(9.,7.1)(0.,){Sine}{0}
\FAProp(9.,12.95)(5.,15.)(0.,){Straight}{0}
\FAProp(9.,12.95)(14.5,10.)(0.,){Straight}{0}
\FAProp(5.,5.)(9.,12.95)(0.,){Sine}{0}
\FAProp(9.,7.1)(5.,5.)(0.,){Straight}{0}
\FAProp(9.,7.1)(14.5,10.)(0.,){Straight}{0}
\FALabel(16.5,14.)[]{\scriptsize $[ (k_1+k_2)^2]$}
\FALabel(2.5,16.77)[b]{\small $p_1$}
\FALabel(2.5,6.77)[b]{\small $p_2$}
\FALabel(10.,0.)[]{$\top{16}(s)$}
\end{feynartspicture}
\hspace{0.3cm}
%
\begin{feynartspicture}(300,300)(1,1)
\FADiagram{}
\FAProp(0.,15.)(5.,15.)(0.,){Straight}{0}
\FAProp(0.,5.)(5.,5.)(0.,){Straight}{0}
\FAProp(20.,10.)(14.5,10.)(0.,){Straight}{0}
\FAProp(5.,15.)(9.,7.1)(0.,){Sine}{0}
\FAProp(9.,12.95)(5.,15.)(0.,){Straight}{0}
\FAProp(9.,12.95)(14.5,10.)(0.,){Straight}{0}
\FAProp(5.,5.)(9.,12.95)(0.,){Sine}{0}
\FAProp(9.,7.1)(5.,5.)(0.,){Straight}{0}
\FAProp(9.,7.1)(14.5,10.)(0.,){Straight}{0}
\FALabel(2.5,16.77)[b]{\small $p_1$}
\FALabel(2.5,6.77)[b]{\small $p_2$}
\FALabel(10.,0.)[]{$\top{17}(s)$}
\end{feynartspicture}

%% file: xbox2L.tex
\FIGURE[t]{
\input{PICTURES/Box2C/Diagrams}
\caption{Non-planar two-loop diagram  with massless internal propagators, and massless external particles.
The internal  momenta shown in the diagram are oriented according to  the arrows. All the external momenta are incoming.}
\label{Fig:Diagrams3}
}

\section{Two-Loop non-planar Box}
\label{Sec:xbox}
The evaluation of the two-loop non-planar box diagram in Fig.~\ref{Fig:Diagrams3}, contributing
to the $2 \to 2$ scattering among massless particles, has already been 
considered in the literature \cite{Tausk:1999vh,Anastasiou:2000mf}. Recently, for its planar
partner, a set of MI's with homogeneous transcendentality was presented
in Ref.~\cite{Henn:2013pwa}.
Our method can be easily applied to it, but instead of showing the case
of the ladder-box diagram, in this section, we compute the additional MI's
required for determining the non-planar contribution, having
expressions with manifest homogeneous transcendentality as well.

The integrals, in this case, are functions of the invariants
$s = (p_1+p_2)^2$,
$t = (p_1+p_3)^2$, and
$u = (p_2+p_3)^2$, with $p_i^2=0$, and $s+t+u=0$. 

\noindent
We adopt the following initial choice of MI's,
\begin{align}
&
\begin{aligned}
f_1 &= \eps^2\, s\, \top{a}(s)\,,   &
f_2 &=  \eps^2\, t\, \top{a}(t)\,, & 
f_3 &=  \eps^2\, u\, \top{a}(u)\,,   \nn\allowdisplaybreaks[1]
f_4 &=  \eps^3\, s\, \top{b}(s)\,,  &
f_5 &=  \eps^3\, s\, t\, \top{c}(s,t) \,, &  
f_6 &=  \eps^3\, s\, u\, \top{c}(s,u) \,,   \nn\allowdisplaybreaks[1]
f_7 &=  \eps^4\, u\, \top{d}(s,t) \,, &
f_8 &= \eps^4\, s\, \top{d}(t,u) \,,  &
f_9 &=  \eps^4\, t\, \top{d}(u,s)  \,,  \nn\allowdisplaybreaks[1]
f_{10} &= \eps^4\, s^2\, \top{e}(s)\,, &&&&  \nonumber
\end{aligned}  \nn\allowdisplaybreaks[1]
f_{11} = {} &         \eps^4\, s\, t\, u\, \top{f}(s,t)
                -\frac{3}{4\, s\, (4 \eps+1)} \left[    \eps^2 \left(s^2\, \top{a}(s) + t^2\, \top{a}(t) + u^2\, \top{a}(u) \,\right) \right. \nn\allowdisplaybreaks[1]
      &          \left. -\,4 \eps^4 \left(u^2\, \top{d}(s,t) + s^2\, \top{d}(t,u) + t^2\, \top{d}(u,s) \right)    \right]       \, ,          \nn\allowdisplaybreaks[1]
 f_{12} = {} &      \eps^4\, s\, t\, \top{g}(s,t)
                -\frac{3}{8\, u\, (4 \eps+1)} \left[    \eps^2 \left(s^2\, \top{a}(s) + t^2\, \top{a}(t) + u^2\, \top{a}(u) \,\right) \right. \nn\allowdisplaybreaks[1]
      &          \left. -\,4 \eps^4 \left(u^2\, \top{d}(s,t) + s^2\, \top{d}(t,u) + t^2\, \top{d}(u,s) \right)    \right] \, ,
\end{align} 
where the integrals $\top{i}$ correspond to the diagrams in
Fig.~\ref{Fig:Masters3}.
We notice that the integrals $f_1, \ldots, f_9$  are common to the two-loop planar box
diagram~\cite{Henn:2013pwa}.
The set $f$ of MI's obeys a system of differential equations the variable $x$, defined as, 
\bea
x = - {t\over s} \ ,
\eea
which is linear in $\eps$. 
According to the procedure in Section~\ref{Sec:diffeq}, we can
build the matrix $B_0(x)$ ruling the change of basis $f(\eps,x) =
B_0(x) g(\eps,x)$, so that the new MI's,
\begin{align}
g_i = {} & f_i \ ,  \qquad  i=1,\ldots, 10\,  ,  \nn\allowdisplaybreaks[1]
g_{11}  = {} & \frac{s}{8\, t\, u}   \,  \big[ 3 f_1 (3\, t-5\, u)-3 f_2 (t+4\, u)+3 f_3 (2\,t+u)-16 f_5 \,u+8 f_6 \,t \nn\allowdisplaybreaks[1] 
       &  -60 f_ 7 \,u-12 f_ 8 (t-u)+36 f_ 9 \,t-8 f_{11}\, u-8 f_{12}\, u \big] \ , \nn\allowdisplaybreaks[1]
g_{12} = {} & \frac{s}{8\, u} \,  \left( 9 f_1-3 f_2+6 f_3+8 f_6-12 f_8+36 f_9 \right) +f_{12} \, ,
\label{Eq:MIfinBox}
\end{align}
obey the canonical system,
\begin{equation}
\partial_x g(\eps,x)=\epsilon \hat{A}_1(x) g(\eps,x) \ , \qquad \hat{A}(x)= \frac{M_1}{x}+ \frac{M_2}{1-x} \, ,
\end{equation}
with
\FIGURE[t]{
\input{PICTURES/Box2C/Masters}

\caption{MI's for the two-loop diagram in Fig.~\ref{Fig:Diagrams3}. All the external 
momenta depicted are incoming. In the last integral the loop momenta have to be 
fixed according to Fig.~\ref{Fig:Diagrams3} and a term 
$(k_2+p_1)^2$ enters the numerator of its integrand. A dot indicates a squared propagator.}
\label{Fig:Masters3}
}
\begin{align}
M_1 &=  \small
\left(
\begin{array}{cccccccccccc}
 0 & 0 & 0 & 0 & 0 & 0 & 0 & 0 & 0 & 0 & 0 & 0 \\
 0 & -2 & 0 & 0 & 0 & 0 & 0 & 0 & 0 & 0 & 0 & 0 \\
 0 & 0 & 0 & 0 & 0 & 0 & 0 & 0 & 0 & 0 & 0 & 0 \\
 0 & 0 & 0 & 0 & 0 & 0 & 0 & 0 & 0 & 0 & 0 & 0 \\
 0 & -\frac{3}{2} & 0 & 0 & -2 & 0 & 0 & 0 & 0 & 0 & 0 & 0 \\
 0 & 0 & \frac{3}{2} & -3 & 0 & 1 & 0 & 0 & 0 & 0 & 0 & 0 \\
 -\frac{1}{2} & \frac{1}{2} & 0 & 0 & 0 & 0 & -2 & 0 & 0 & 0 & 0 & 0 \\
 0 & \frac{1}{2} & -\frac{1}{2} & 0 & 0 & 0 & 0 & -2 & 0 & 0 & 0 & 0 \\
 0 & 0 & 0 & 0 & 0 & 0 & 0 & 0 & 2 & 0 & 0 & 0 \\
 0 & 0 & 0 & 0 & 0 & 0 & 0 & 0 & 0 & 0 & 0 & 0 \\
 -6 & -6 & -\frac{9}{2} & 0 & -4 & -2 & -18 & -12 & -12 & 1 & 1 & -2 \\
 \frac{3}{4} & \frac{9}{4} & -\frac{21}{4} & 3 & 2 & -3 & 12 & -6 & -18 & 0 & 0 & -2 \\
\end{array}
\right) 
\ ,  \normalsize\nn  \allowdisplaybreaks[1] 
M_2  &=  \small  \left(
\begin{array}{cccccccccccc}
 0 & 0 & 0 & 0 & 0 & 0 & 0 & 0 & 0 & 0 & 0 & 0 \\
 0 & 0 & 0 & 0 & 0 & 0 & 0 & 0 & 0 & 0 & 0 & 0 \\
 0 & 0 & 2 & 0 & 0 & 0 & 0 & 0 & 0 & 0 & 0 & 0 \\
 0 & 0 & 0 & 0 & 0 & 0 & 0 & 0 & 0 & 0 & 0 & 0 \\
 0 & -\frac{3}{2} & 0 & 3 & -1 & 0 & 0 & 0 & 0 & 0 & 0 & 0 \\
 0 & 0 & \frac{3}{2} & 0 & 0 & 2 & 0 & 0 & 0 & 0 & 0 & 0 \\
 0 & 0 & 0 & 0 & 0 & 0 & -2 & 0 & 0 & 0 & 0 & 0 \\
 0 & \frac{1}{2} & -\frac{1}{2} & 0 & 0 & 0 & 0 & 2 & 0 & 0 & 0 & 0 \\
 \frac{1}{2} & 0 & -\frac{1}{2} & 0 & 0 & 0 & 0 & 0 & 2 & 0 & 0 & 0 \\
 0 & 0 & 0 & 0 & 0 & 0 & 0 & 0 & 0 & 0 & 0 & 0 \\
 -6 & -6 & -\frac{9}{2} & 0 & -4 & -2 & -18 & -12 & -12 & 1 & 1 & -2 \\
 -\frac{21}{4} & \frac{9}{4} & -\frac{27}{4} & -6 & 2 & -4 & 12 & -6 & -24 & 1 & -1 & 0 \\
\end{array}
\right) \, \normalsize.%
\end{align}
The solution of the system can be expressed as Dyson series, as well as
Magnus series, in terms of one-dimensional HPL's~\cite{Remiddi:1999ew}. All MI's have been
computed in the scattering kinematics, i.e. $s>0$, $t<0$, $u<0$ with $|s|>|t|$, which gives $0 < x < 1$. As long as the planar sub
topologies are concerned, one can fix the boundary conditions using the
regularity properties of the integrals in some special kinematical points.
On the other hand, the analyticity structure of the crossed box is more complicated, since it involves at the same time 
cuts in all three Mandelstam variables $s$, $t$, $u$. 
 Nevertheless, in this particular case, the boundaries can be fixed 
by direct comparison with the results presented in~\cite{Tausk:1999vh,Anastasiou:2000mf}.
The expressions of the transcendentally homogeneous MI's $g$ are shown
in Appendix~\ref{results:xbox2L}, and collected in the ancillary file {\tt <xbox2L.m>}.

%% file: PICTURES/Box2C/Diagrams.tex
\unitlength=0.3bp%
\hspace{5cm}
\begin{feynartspicture}(300,300)(1,1)
\FADiagram{}
\FAProp(0.,15.)(5.5,13.5)(0.,){Straight}{0}
\FAProp(0.,5.)(5.5,6.5)(0.,){Straight}{0}
\FAProp(20.,15.)(14.5,13.5)(0.,){Straight}{0}
\FAProp(20.,5.)(14.5,6.5)(0.,){Straight}{0}
\FAProp(5.5,13.5)(5.5,6.5)(0.,){Straight}{-1}
\FALabel(3.7,10.)[]{\scriptsize $k_1$}
\FAProp(5.5,13.5)(14.5,13.5)(0.,){Straight}{0}
\FAProp(5.5,6.5)(14.5,6.5)(0.,){Straight}{0}
%
\FAProp(14.5,13.5)(12.25, 10)(0.,){Straight}{-1}
\FAProp(12.25, 10)(10.,6.5)(0.,){Straight}{0}
\FAProp(14.5,6.5)(10.,13.5)(0.,){Straight}{0}
\FALabel(19.8, 11.75)[]{\scriptsize $k_1-k_2-p_2$}
\FALabel(4.2,16.4)[]{\small $p_1$}
\FALabel(4.2,3.6)[]{\small $p_2$}
\FALabel(15.8,16.4)[]{\small $p_3$}
\FALabel(15.8,3.6)[]{\small $p_4$}
\end{feynartspicture}
\hspace{5cm}

%% file: PICTURES/Box2C/Masters.tex
\unitlength=0.25bp%
\hspace{0.3cm}
\begin{feynartspicture}(300,300)(1,1)
\FADiagram{}
\FAProp(0.,10.)(6.,10.)(0.,){Straight}{0}
\FAProp(20.,10.)(14.,10.)(0.,){Straight}{0}
\FAProp(14.,10.)(6.,10.)(-1.,){Straight}{0}
\FAProp(14.,10.)(6.,10.)(0.,){Straight}{0}
\FAVert(10.,10.){0}
\FAVert(10.,6.){0}
\FAProp(14.,10.)(6.,10.)(1.,){Straight}{0}
\FALabel(2.,8.93)[t]{\small $p_{12}$}
\FALabel(10.,0.)[]{$\top{a}(s)$}
\end{feynartspicture}
\hspace{0.3cm}
%
\begin{feynartspicture}(300,300)(1,1)
\FADiagram{}
\FAProp(0.,15.)(6.,15.)(0.,){Straight}{0}
\FAProp(0.,5.)(6.,5.)(0.,){Straight}{0}
\FAProp(20.,10.)(14.5,10.)(0.,){Straight}{0}
\FAProp(14.5,10.)(6.,5.)(0.,){Straight}{0}
\FAProp(14.5,10.)(6.,15.)(0.,){Straight}{0}
\FAProp(6.,5.)(6.,15.)(-0.4,){Straight}{0}
\FAProp(6.,5.)(6.,15.)(0.4,){Straight}{0}
\FAVert(8.,10.){0}
\FALabel(2.5,16.77)[b]{\small $p_1$}
\FALabel(2.5,6.77)[b]{\small $p_2$}
\FALabel(10.,0.)[]{$\top{b}(s)$}
\end{feynartspicture}
%
\begin{feynartspicture}(300,300)(1,1)
\FADiagram{}
\FAProp(0.,15.)(6.5,13.5)(0.,){Straight}{0}
\FAProp(0.,5.)(6.5,6.5)(0.,){Straight}{0}
\FAProp(20.,15.)(13.5,13.5)(0.,){Straight}{0}
\FAProp(20.,5.)(13.5,6.5)(0.,){Straight}{0}
\FAProp(6.5,13.5)(6.5,6.5)(0.,){Straight}{0}
\FAProp(6.5,13.5)(13.5,13.5)(0.,){Straight}{0}
\FAProp(6.5,6.5)(13.5,6.5)(0.,){Straight}{0}
\FAProp(13.5,13.5)(13.5,6.5)(0.4,){Straight}{0}
\FAProp(13.5,13.5)(13.5,6.5)(-0.4,){Straight}{0}
\FAVert(14.9,10.0){0}
\FALabel(4.2,16.4)[]{\small $p_1$}
\FALabel(4.2,3.6)[]{\small $p_2$}
\FALabel(15.8,16.4)[]{\small $p_3$}
\FALabel(15.8,3.6)[]{\small $p_4$}
\FALabel(10.,0.)[]{$\top{c}(s,t)$}
\end{feynartspicture}
\hspace{0.3cm}
%
\begin{feynartspicture}(300,300)(1,1)
\FADiagram{}
\FAProp(0.,15.)(6.5,13.5)(0.,){Straight}{0}
\FAProp(0.,5.)(6.5,6.5)(0.,){Straight}{0}
\FAProp(20.,15.)(13.5,13.5)(0.,){Straight}{0}
\FAProp(20.,5.)(13.5,6.5)(0.,){Straight}{0}
\FAProp(6.5,13.5)(6.5,6.5)(0.,){Straight}{0}
\FAProp(6.5,13.5)(13.5,13.5)(0.,){Straight}{0}
\FAProp(6.5,6.5)(13.5,6.5)(0.,){Straight}{0}
\FAProp(13.5,13.5)(13.5,6.5)(0.,){Straight}{0}
\FAProp(6.5,6.5)(13.5,13.5)(0.,){Straight}{0}
\FALabel(4.2,16.4)[]{\small $p_1$}
\FALabel(4.2,3.6)[]{\small $p_2$}
\FALabel(15.8,16.4)[]{\small $p_3$}
\FALabel(15.8,3.6)[]{\small $p_4$}
\FALabel(10.,0.)[]{$\top{d}(s,t)$}
\end{feynartspicture}
\hspace{0.3cm}
%
\begin{feynartspicture}(300,300)(1,1)
\FADiagram{}
\FAProp(0.,15.)(5.,15.)(0.,){Straight}{0}
\FAProp(0.,5.)(5.,5.)(0.,){Straight}{0}
\FAProp(20.,10.)(14.5,10.)(0.,){Straight}{0}
\FAProp(5.,15.)(9.,7.1)(0.,){Straight}{0}
\FAProp(9.,12.95)(5.,15.)(0.,){Straight}{0}
\FAProp(9.,12.95)(14.5,10.)(0.,){Straight}{0}
\FAProp(5.,5.)(9.,12.95)(0.,){Straight}{0}
\FAProp(9.,7.1)(5.,5.)(0.,){Straight}{0}
\FAProp(9.,7.1)(14.5,10.)(0.,){Straight}{0}
\FALabel(2.5,16.77)[b]{\small $p_1$}
\FALabel(2.5,6.77)[b]{\small $p_2$}
\FALabel(10.,0.)[]{$\top{e}(s)$}
\end{feynartspicture}
\hspace{0.3cm}
%
\begin{feynartspicture}(300,300)(1,1)
\FADiagram{}
\FAProp(0.,15.)(5.5,13.5)(0.,){Straight}{0}
\FAProp(0.,5.)(5.5,6.5)(0.,){Straight}{0}
\FAProp(20.,15.)(14.5,13.5)(0.,){Straight}{0}
\FAProp(20.,5.)(14.5,6.5)(0.,){Straight}{0}
\FAProp(5.5,13.5)(5.5,6.5)(0.,){Straight}{0}
\FAProp(5.5,13.5)(14.5,13.5)(0.,){Straight}{0}
\FAProp(5.5,6.5)(14.5,6.5)(0.,){Straight}{0}
%
\FAProp(14.5,13.5)(10.,6.5)(0.,){Straight}{0}
\FAProp(14.5,6.5)(10.,13.5)(0.,){Straight}{0}
\FALabel(4.2,16.4)[]{\small $p_1$}
\FALabel(4.2,3.6)[]{\small $p_2$}
\FALabel(15.8,16.4)[]{\small $p_3$}
\FALabel(15.8,3.6)[]{\small $p_4$}
\FALabel(10.,0.)[]{$\top{f}(s,t)$}
\end{feynartspicture}
\hspace{0.3cm}
%
\begin{feynartspicture}(300,300)(1,1)
\FADiagram{}
\FAProp(0.,15.)(5.5,13.5)(0.,){Straight}{0}
\FAProp(0.,5.)(5.5,6.5)(0.,){Straight}{0}
\FAProp(20.,15.)(14.5,13.5)(0.,){Straight}{0}
\FAProp(20.,5.)(14.5,6.5)(0.,){Straight}{0}
\FAProp(5.5,13.5)(5.5,6.5)(0.,){Straight}{0}
\FAProp(5.5,13.5)(14.5,13.5)(0.,){Straight}{0}
\FAProp(5.5,6.5)(14.5,6.5)(0.,){Straight}{0}
%
\FAProp(14.5,13.5)(10.,6.5)(0.,){Straight}{0}
\FAProp(14.5,6.5)(10.,13.5)(0.,){Straight}{0}
\FALabel(4.2,16.4)[]{\small $p_1$}
\FALabel(4.2,3.6)[]{\small $p_2$}
\FALabel(15.8,16.4)[]{\small $p_3$}
\FALabel(15.8,3.6)[]{\small $p_4$}
\FALabel(22.,10.)[]{\scriptsize $[(k_2+p_1)^2]$}
\FALabel(10.,0.)[]{$\top{g}(s,t)$}
\end{feynartspicture}
\hspace{0.3cm}
%
%
%
%
%
%

%% file: polyeps.tex
\section{Polynomial $\epsilon$ dependence}
The cases discussed above admitted an initial choice of MI's $f$
obeying a system of differential equations linear in $\eps$. 
We cannot be sure that this feature is general, and holds for any
scattering process in dimensional regularization.
Nevertheless, the use of Magnus series enables us to generalize our
algorithm to the case of systems of DE's whose matrix is a
{\it polynomial} in
$\epsilon$.
In fact, let us consider a system of equations where $A$ is of degree
$\kappa$ in $\eps$,
\bea
\partial_x f(\epsilon,x) = A(\epsilon,x) \,  f(\epsilon,x)  \ , \qquad
A(\eps, x ) \equiv \sum_{k=0}^{\kappa} \eps^k { A}_k(x) \ .
\label{Eq:DiffEqPol}
\eea
By iterating the algorithm described in Section~\ref{Sec:diffeq},
the solution of the differential equation~(\ref{Eq:DiffEqPol}) can be expressed in terms of 
a chain of products of Magnus exponentials,
\bea
f(\epsilon,x) = 
B_0(x)  
B_1(\eps, x)  
\cdots  
B_\kappa(\eps, x)   
\ef{\kappa}(\eps) \ ,  \qquad 
B_k(\eps,x) \equiv e^{\magnus{\epsilon^k \hat A_k}(x,x_0)} \ , 
\label{eq:polycase:sol}
\eea
where the kernel  $\hat A_k$ is defined as 
\bea
 {\hat { A}}_k(\eps, x)  &=&  {\hat { A}}_k^{(k)}(\eps, x)  \, , \nn
 {\hat { A}}_k^{(j)}(\eps, x) &=&  
B_{j-1}^{-1}(\eps,x)  \cdots
B_{1}^{-1}(\eps,x) 
B_{0}^{-1}(x) 
\; { A}_k(x)  \; B_{0}(x) B_{1}(\eps,x) 
 \cdots
 B_{j-1}(\eps,x) \ .  \qquad
\label{eq:polycase:map}
\eea
It is worth to observe that, within our construction, the solution $f$ is given by
repeated transformations. Starting from 
\bea
f(\eps,x) =  B_0(x)\ef{0}(\eps,x) \, , 
\eea
we iteratively write 
$\ef{k}$ as,
\bea
f_{k}(\eps,x) = B_{k+1}(\eps,x)f_{k+1}(\eps,x) \ , \qquad (0 \le k \le \kappa-1) \, ,
\eea
which obeys the system
\bea
\partial_x  \ef{k}(\eps, x) &=& 
\eps^{k}  
\bigg ( \, 
\sum_{j=1}^{\kappa - k} 
\eps^{j} {\hat { A}}_{k+j}^{(k+1)}(\eps,x)  \, 
\bigg  )
\ \ef{k}(\eps, x)
 \ .
\eea
The generalization of the canonical system Eq.~(\ref{eq:cansys}) is obtained at the last
step of the iteration, when $k=\kappa-1$, 
\bea
\ef{\kappa-1}(\eps,x) = B_\kappa(\eps,x)\ef{\kappa}(\eps) \ , \qquad
\partial_x  \ef{\kappa-1}(\eps, x) &=& \eps^{\kappa}  
{\hat { A}}_{\kappa}(\eps,x) 
\ \ef{\kappa-1}(\eps, x) \ .
\eea

It is important to remark that the complete factorization of $\epsilon$ is
achieved only if $\kappa=1$, i.e. if the system is linear in $\eps$, because, although 
${\hat { A}}_1$ is independent of $\eps$, ${\hat { A}}_k$ acquires a dependence on
$\eps$  for $k> 1$, {\it cfr.}   Eq.~(\ref{eq:polycase:map}). \\ 

The algorithm here described has a wide range of applicability and  can be used to compute generic sets of MI's,  
provided that the matrix associated to the system of DE's can be Taylor expanded  around $\eps =0$. In this case, 
the MI's are obtained perturbatively by  truncating the $\eps$
expansion of the  matrices associated to the systems of DE's.

%% file: conclusions.tex
\section{Conclusions}
In this article we elaborated on the method of differential
equations for Feynman integrals within the $D$-dimensional
regularization scheme.

The freedom in the choice of the MI's allowed us to 
analyze the paradigmatic case of systems of 
differential equations whose matrix is linear  
in the dimensional parameter, $\epsilon = (4 - D)/2$.
We showed that these systems admit a {\it canonical} form,
where the dependence on $\epsilon$ is factorized from the
kinematic variables, as recently suggested by Henn.

We used  Magnus series  to  obtain the matrix
implementing the transformation from the linear to the canonical
form. The solution of the canonical system is obtained by using   either Dyson
series or Magnus series.  Both series require multiple integrations
which allow one to naturally express the MI's  in
terms of polylogarithms and of  their generalization.

We demonstrated that
the one-loop Bhabha scattering,
the two-loop electron form factors in QED and
the two-loop $2 \to 2$ massless scattering
exhibit a basis of MI's leading to linear systems of DE's.
We then obtained the corresponding canonical bases, in terms of
uniform transcendentality functions.

Finally, we have shown that our procedure can be extended to the more
general case of systems of DE's that are polynomial in $\epsilon$.   

The  range of applicability  of the algorithm is rather  wide and  can be used to compute generic sets of MI's,  
provided that the matrix associated to the system of DE's can be Taylor expanded  around $\eps =0$.

%% file: appendix.tex
\newtheorem{lemma}{Lemma}[section]
\newtheorem{theo}{Theorem}[section]

\section{Magnus Theorem}
\label{App:Magnus}

We closely follow the discussion of Ref.\cite{PolySdeMany}.
Given  an operator, $\Omega$, 
we define the 
derivative of $\Omega^k$ w.r.t. $\Omega$ by its 
action on a generic operator $H$:
\bea
\bigg(
{d \over d\Omega} \Omega^k 
\bigg) H \equiv H \Omega^{k-1} + \Omega H \Omega^{k-2} + \ldots +
\Omega^{k-1} H \, .
\label{Eq:DefDer}
\eea
This definition guarantees that,  when $\Omega=\Omega(x)$ and  $H=\partial_x\Omega$,
 \bea
\partial_x \Omega^k    = \bigg( {d \over d\Omega} \Omega^k   \bigg) \,  \partial_x\Omega.
\label{Eq:ProDer}
\eea
The definition~(\ref{Eq:DefDer}) reduces to  $k H \Omega^{k-1}$ when $[\Omega, H] = 0$, therefore  it is 
natural to write it as $k H \Omega^{k-1}$ plus correction terms involving (iterated) commutators. Using the 
adjoint operator 
\begin{equation}
\mbox{ad}_\Omega (H) \equiv [\Omega, H], 
\end{equation}
and its iterated application  $\mbox{ad}^{\; i}_\Omega$ we obtain
\bea
\bigg(
{d \over d\Omega} \Omega^2 
\bigg) H  &=&  H\Omega + \Omega H = 2 H \Omega +\mbox{ad}_\Omega (H) \nn
\bigg(
{d \over d\Omega} \Omega^3 
\bigg) H  &=&  H\Omega^2 + \Omega H \Omega  + \Omega^2 H = 3 H \Omega^2 + 3 [\Omega,H]\Omega + \mbox{ad}^{\, 2}_\Omega (H)
 \nn [0.5ex]
\vdots \qquad    && \qquad \vdots \nn  [0.5ex]
\bigg(
{d \over d\Omega} \Omega^k
\bigg) H  &=& \sum_{i=0}^{k-1}
\begin{pmatrix}
k \\
i+1
\end{pmatrix}
 \mbox{ad}^{\, i}_\Omega (H) \ \Omega^{k-i-1}\, .
\eea
The last equation can be obtained by  induction using the relation 
\bea
\Omega \; \mbox{ad}^{\, i}_\Omega (H)    =   \mbox{ad}^{\, i}_\Omega (H) \;  \Omega +  \mbox{ad}^{\, i+1}_\Omega (H) 
\eea
\bigskip

\noindent
The exponential of a matrix $\Omega$ is defined via a series expansion:
\bea 
e^\Omega \equiv \sum_{k \ge 0} {1 \over k!} \Omega^k \ .
\label{Eq:DefExp}
\eea
The derivative and the inverse of the exponential of a matrix can be straightforwardly obtained
by using the previous results:

\begin{lemma}[Derivative of the exponential]
The derivative of the matrix exponential can be derived from its
action on a generic operator $H$ and reads as follows
\label{lemma1}
\end{lemma}
\bea
\bigg(
{d \over d\Omega} e^\Omega\bigg) H 
= d \exp_\Omega(H)  \, e^\Omega \ ,  \qquad 
d \exp_\Omega(H) 
\equiv  
\sum_{k \ge 0} {1 \over (k+1)!} \; \mbox{ad}^{\, i}_\Omega (H) 
\eea

\begin{lemma}[Inverse of the exponential]
If the eigenvalues of $\mbox{ad}_\Omega$
are different from $2 \ell \pi i $ with $\ell \in \{ \pm 1, \pm 2, \ldots \}$, then $d \exp_\Omega$
is invertible, and 
\label{lemma2}
\end{lemma}
\bea
d \exp_\Omega^{-1}(H) 
=
\sum_{k \ge 0} {\beta_k \over k!}  \; \mbox{ad}^{\, i}_\Omega (H)   \ ,
\eea
where $\beta_k$ are the Bernoulli numbers,  whose generating function is
\bea
{t \over e^t-1} = \sum_{k=0}^\infty \ {\beta_k \over k!} \ t^k \ .
\eea
\bigskip

\noindent
We have now all the ingredients to prove the  following~\cite{Magnus}
\begin{theo}[Magnus]
The solution of  a generic linear matrix differential equation
\bea
\partial_x Y = A(x) Y \ , \quad Y(x_0) = Y_0 \,
\eea
can be written as
\bea
Y(x) = e^{\Omega(x,x_0)} \ Y(x_0) \equiv  e^{\Omega(x)} \ Y_0  \, 
\label{Eq:solMagnus}
\eea
where $\Omega(x)$ can be computed by solving the differential equation,
\bea
\partial_x \Omega = d \exp_\Omega^{-1} \Big( A(x) \Big)  \ ,
\qquad \Omega(x_0) = 0 \ .
\label{Eq:Thesis}
\eea
\label{theoMagnus}
\end{theo}
{\bf Proof} \quad  Let us
consider the derivative of~(\ref{Eq:solMagnus}). Using the definition~(\ref{Eq:DefExp})
and the property~(\ref{Eq:ProDer}) we have 
\bea
\partial_x Y 
= 
\bigg({d \over d \Omega} e^\Omega \bigg) \ 
\partial_x \Omega \ Y_0 
= 
d \exp_\Omega ( \partial_x  \Omega)  \ e^\Omega \ Y_0 
=  d \exp_\Omega ( \partial_x  \Omega )  Y(x) \, . \nonumber 
\eea
The {\it r.h.s.} is equal to $A(x) Y(x)$ when
\bea
 d \exp_\Omega (  \partial_x   \Omega ) = A(x) \ . 
\label{eq:dexpOmega}
\eea
The relation~(\ref{Eq:Thesis}) is thus proven by applying the operator $d
\exp_\Omega^{-1} $ to both sides of~(\ref{eq:dexpOmega}). \; \hfill $\Box$
\bigskip

\noindent
The differential equation for $\Omega$ explicitly reads,
\bea
\partial_x \Omega = A(x) - {1\over 2} [\Omega, A(x)] 
+ {1 \over 12} [\Omega,[\Omega, A(x)] ] 
+ \ldots \ ,
\eea
and the solution can be written as a series, called  {\it Magnus expansion},
\bea
\Omega=\sum_{n=1}^\infty \Omega_n(x) \ ,   \qquad 
\Omega_n(x)  &=& \sum_{j=1}^{n-1} {\beta_j \over j !} \int_{x_0}^x
S_n^{(j)}(\tau) d\tau  \, .
\eea
The coefficients $\beta_j$ are the  Bernoulli numbers while the  integrands $S_n^{(j)}$ can be computed recursively,
\bea
S_n^{(1)} &=&  [\Omega_{n-1},A] \ , \nn
S_n^{(j)} &=& \sum_{m=j-1}^{n-j} \,  \left [ \Omega_m, S_{n-m}^{(j-1)} \right ] \ 
\qquad 2 \le j \le n-2 \ , \nn
S_n^{(n-1)} &=& [\Omega_1, A] \ .
\eea

%% file: MIvertex.tex
\section{Master Integrals for the two-loop QED vertices}
\label{results:vertex2L}
In this Appendix we collect   the  $17$ MI's of the 
two-loop QED vertices introduced  in Eq.~(\ref{Eq:MIfinVertex}). In Section~\ref{Sec:vertex}, we have obtained them starting from the integrals $\top{i}$ depicted 
in Fig.~\ref{Fig:Masters2}, which are normalized according to the integration measure (Minkowskian metric is understood)
\begin{displaymath}
\left ( \frac{m^{2 \eps}}{\Gamma(1+\eps)} \right )^2 \, \int \,
\frac{d^D k_1}{\pi^{D/2}} \, \int \,   \frac{d^D k_2 }{\pi^{D/2}} \, .
\end{displaymath}
The MI's exhibit uniform transcendentality. In the following  we present the expression of the coefficients
of their  expansion around $\eps=0$ up to $\mathcal{O}(\eps^4)$.
The coefficients  $g_{i}^{(a)}$ are defined as follows:
\begin{displaymath}
g_i = \sum_{a=0}^{4} \eps^a \ g_{i}^{(a)} \ , \qquad i=1,\ldots,17  \, .
\end{displaymath}

\begin{subequations}
\begin{align}
                   g_{1}^{(0)}   = {} & 
                         -1\,,\\ \allowdisplaybreaks[1] 
                   g_{1}^{(1)}   = {} & 
                          0 \,,\\ \allowdisplaybreaks[1] 
                   g_{1}^{(2)}   = {} & 
                          0 \,,\\ \allowdisplaybreaks[1] 
                   g_{1}^{(3)}   = {} & 
                          0 \,,\\ \allowdisplaybreaks[1] 
                   g_{1}^{(4)}   = {} & 
                          0 \,,
\end{align}
\end{subequations}

\begin{subequations}
\begin{align}
                   g_{2}^{(0)}   = {} & 
                          0 \,,\\ \allowdisplaybreaks[1] 
                   g_{2}^{(1)}   = {} & 
                         -\,\HPL(0; x)\,,\\ \allowdisplaybreaks[1] 
                   g_{2}^{(2)}   = {} & 
                         2 \,\HPL(-1,0; x)-\,\HPL(0,0; x) + \zeta_2\,,\\ \allowdisplaybreaks[1] 
                   g_{2}^{(3)}   = {} & 
                         -4 \,\HPL(-1,-1,0; x)+2 \,\HPL(-1,0,0; x)+2 \,\HPL(0,-1,0; x)\nn \allowdisplaybreaks[1]
                         &-\,\HPL(0,0,0; x) + \,\zeta_2 (\,\HPL(0; x)-2 \,\HPL(-1; x)) + 2 \,\zeta_3\,,\\ \allowdisplaybreaks[1] 
                   g_{2}^{(4)}   = {} & 
                         8 \,\HPL(-1,-1,-1,0; x)-4 \,\HPL(-1,-1,0,0; x)-4 \,\HPL(-1,0,-1,0; x)\nn \allowdisplaybreaks[1]
                         &  +2 \,\HPL(-1,0,0,0; x)-4 \,\HPL(0,-1,-1,0; x)+2 \,\HPL(0,-1,0,0; x)\nn \allowdisplaybreaks[1]
                         & +2 \,\HPL(0,0,-1,0; x)-\,\HPL(0,0,0,0; x) 
                         + \zeta_2 (4 \,\HPL(-1,-1; x) \nn \allowdisplaybreaks[1]
                         & -2 \,\HPL(-1,0; x)-2 \,\HPL(0,-1; x)+\,\HPL(0,0; x))  \nn \allowdisplaybreaks[1]
                         & -2 \,\zeta_3 (2 \,\HPL(-1; x)-\,\HPL(0; x)) + \frac{9 \,\zeta _4}{4}\,,
\end{align}
\end{subequations}

\begin{subequations}
\begin{align}
                   g_{3}^{(0)}   = {} & 
                          0 \,,\\ \allowdisplaybreaks[1] 
                   g_{3}^{(1)}   = {} & 
                          0 \,,\\ \allowdisplaybreaks[1] 
                   g_{3}^{(2)}   = {} & 
                         -2 \,\HPL(0,0; x)\,,\\ \allowdisplaybreaks[1] 
                   g_{3}^{(3)}   = {} & 
                         8 \,\HPL(-1,0,0; x)+4 \,\HPL(0,-1,0; x)-6 \,\HPL(0,0,0; x) + 2 \,\zeta_2 \,\HPL(0; x)\,,\\ \allowdisplaybreaks[1] 
                   g_{3}^{(4)}   = {} & 
                         -32 \,\HPL(-1,-1,0,0; x)-16 \,\HPL(-1,0,-1,0; x)+24 \,\HPL(-1,0,0,0; x) \nn \allowdisplaybreaks[1]
                         &-8 \,\HPL(0,-1,-1,0; x)+20 \,\HPL(0,-1,0,0; x)+12 \,\HPL(0,0,-1,0; x) \nn \allowdisplaybreaks[1]
                         &-14 \,\HPL(0,0,0,0; x) 
                          - 2 \,\zeta_2 (4 \,\HPL(-1,0; x)+2 \,\HPL(0,-1; x)-3 \,\HPL(0,0; x))  \nn \allowdisplaybreaks[1]
                         &+4 \,\zeta_3 \,\HPL(0; x) - \frac{5 \,\zeta _4}{2}\,,
\end{align}
\end{subequations}

\begin{subequations}
\begin{align}
                   g_{4}^{(0)}   = {} & 
                         \frac{1}{4}\,,\\ \allowdisplaybreaks[1] 
                   g_{4}^{(1)}   = {} & 
                          0 \,,\\ \allowdisplaybreaks[1] 
                   g_{4}^{(2)}   = {} & 
                         \zeta_2\,,\\ \allowdisplaybreaks[1] 
                   g_{4}^{(3)}   = {} & 
                         2 \,\zeta_3\,,\\ \allowdisplaybreaks[1] 
                   g_{4}^{(4)}   = {} & 
                         16 \,\zeta _4\,,
\end{align}
\end{subequations}

\begin{subequations}
\begin{align}
                   g_{5}^{(0)}   = {} & 
                          0 \,,\\ \allowdisplaybreaks[1] 
                   g_{5}^{(1)}   = {} & 
                          \HPL(0; x)\,,\\ \allowdisplaybreaks[1] 
                   g_{5}^{(2)}   = {} & 
                         -6 \,\HPL(-1,0; x)+5 \,\HPL(0,0; x)+2 \,\HPL(1,0; x) - \zeta_2\,,\\ \allowdisplaybreaks[1] 
                   g_{5}^{(3)}   = {} & 
                          36 \,\HPL(-1,-1,0; x)-24 \,\HPL(-1,0,0; x)-12 \,\HPL(-1,1,0; x) \nn \allowdisplaybreaks[1]
                         & -30 \,\HPL(0,-1,0; x)+13 \,\HPL(0,0,0; x)+10 \,\HPL(0,1,0; x) \nn \allowdisplaybreaks[1]
                         & -12 \,\HPL(1,-1,0; x)+6 \,\HPL(1,0,0; x)+4 \,\HPL(1,1,0; x) \nn \allowdisplaybreaks[1]
                         & +\zeta_2 (6 \,\HPL(-1; x)-5 \,\HPL(0; x)-2 \,\HPL(1; x)) - 14 \,\zeta_3\,,\\ \allowdisplaybreaks[1] 
                   g_{5}^{(4)}   = {} & 
                          -216 \,\HPL(-1,-1,-1,0; x)+144 \,\HPL(-1,-1,0,0; x) \nn \allowdisplaybreaks[1]
                         &+72 \,\HPL(-1,-1,1,0; x)+144 \,\HPL(-1,0,-1,0; x)-60 \,\HPL(-1,0,0,0; x) \nn \allowdisplaybreaks[1]
                         &-48 \,\HPL(-1,0,1,0; x)+72 \,\HPL(-1,1,-1,0; x)-48 \,\HPL(-1,1,0,0; x) \nn \allowdisplaybreaks[1]
                         &-24 \,\HPL(-1,1,1,0; x)+180 \,\HPL(0,-1,-1,0; x)-120 \,\HPL(0,-1,0,0; x) \nn \allowdisplaybreaks[1]
                         &-60 \,\HPL(0,-1,1,0; x)-78 \,\HPL(0,0,-1,0; x)+29 \,\HPL(0,0,0,0; x) \nn \allowdisplaybreaks[1]
                         &+26 \,\HPL(0,0,1,0; x)-60 \,\HPL(0,1,-1,0; x)+54 \,\HPL(0,1,0,0; x) \nn \allowdisplaybreaks[1]
                         &+20 \,\HPL(0,1,1,0; x)+72 \,\HPL(1,-1,-1,0; x)-48 \,\HPL(1,-1,0,0; x) \nn \allowdisplaybreaks[1]
                         &-24 \,\HPL(1,-1,1,0; x)-36 \,\HPL(1,0,-1,0; x)+14 \,\HPL(1,0,0,0; x) \nn \allowdisplaybreaks[1]
                         &+12 \,\HPL(1,0,1,0; x)-24 \,\HPL(1,1,-1,0; x)+20 \,\HPL(1,1,0,0; x) \nn \allowdisplaybreaks[1]
                         &+8 \,\HPL(1,1,1,0; x) + \zeta_2 (-36 \,\HPL(-1,-1; x)+24 \,\HPL(-1,0; x) \nn \allowdisplaybreaks[1]
                         &+12 \,\HPL(-1,1; x)+30 \,\HPL(0,-1; x)-13 \,\HPL(0,0; x)-10 \,\HPL(0,1; x) \nn \allowdisplaybreaks[1]
                         &+12 \,\HPL(1,-1; x)-6 \,\HPL(1,0; x)-4 \,\HPL(1,1; x)) 
                         + 2 \,\zeta_3 (33 \,\HPL(-1; x) \nn \allowdisplaybreaks[1]
                         &-17 \,\HPL(0; x)-8 \,\HPL(1; x)) - \frac{61 \,\zeta _4}{4}\,,
\end{align}
\end{subequations}

\begin{subequations}
\begin{align}
                   g_{6}^{(0)}   = {} & 
                          0 \,,\\ \allowdisplaybreaks[1] 
                   g_{6}^{(1)}   = {} & 
                          0 \,,\\ \allowdisplaybreaks[1] 
                   g_{6}^{(2)}   = {} & 
                         2 \,\HPL(0,0; x)\,,\\ \allowdisplaybreaks[1] 
                   g_{6}^{(3)}   = {} & 
                         -12 \,\HPL(0,-1,0; x)+6 \,\HPL(0,0,0; x)+4 \,\HPL(0,1,0; x)-4 \,\HPL(1,0,0; x) \nn \allowdisplaybreaks[1]
                         &-2 \,\zeta_2 \,\HPL(0; x) + \nn \allowdisplaybreaks[1]
                         &-6 \,\zeta_3\,,\\ \allowdisplaybreaks[1] 
                   g_{6}^{(4)}   = {} & 
                         72 \,\HPL(0,-1,-1,0; x)-48 \,\HPL(0,-1,0,0; x)-24 \,\HPL(0,-1,1,0; x) \nn \allowdisplaybreaks[1]
                      & -36 \,\HPL(0,0,-1,0; x)+14 \,\HPL(0,0,0,0; x)+12 \,\HPL(0,0,1,0; x) \nn \allowdisplaybreaks[1]
                      & -24 \,\HPL(0,1,-1,0; x)+20 \,\HPL(0,1,0,0; x)+8 \,\HPL(0,1,1,0; x) \nn \allowdisplaybreaks[1]
                      & +24 \,\HPL(1,0,-1,0; x)-12 \,\HPL(1,0,0,0; x)-8 \,\HPL(1,0,1,0; x) \nn \allowdisplaybreaks[1]
                      & +8 \,\HPL(1,1,0,0; x) + 2 \,\zeta_2 (6 \,\HPL(0,-1; x)-3 \,\HPL(0,0; x) \nn \allowdisplaybreaks[1]
                      & -2 \,\HPL(0,1; x)+2 \,\HPL(1,0; x)) - 4 \,\zeta_3 (4 \,\HPL(0; x)-3 \,\HPL(1; x)) 
                         -\frac{13 \,\zeta _4}{2}\,,
\end{align}
\end{subequations}

\begin{subequations}
\begin{align}
                   g_{7}^{(0)}   = {} & 
                          0 \,,\\ \allowdisplaybreaks[1] 
                   g_{7}^{(1)}   = {} & 
                          0 \,,\\ \allowdisplaybreaks[1] 
                   g_{7}^{(2)}   = {} & 
                         \frac{\,\zeta_2}{2}\,,\\ \allowdisplaybreaks[1] 
                   g_{7}^{(3)}   = {} & 
                         -3 \,\zeta_2 \log{2} + \frac{7 \,\zeta_3}{4}\,,\\ \allowdisplaybreaks[1] 
                   g_{7}^{(4)}   = {} & 
                         \frac{1}{2} \left(24 \,\text{Li}_4{\frac12}+\log^4{2}\right) 
                         + 6 \,\zeta_2 \log^2{2} - \frac{31 \,\zeta _4}{4}\,,
\end{align}
\end{subequations}

\begin{subequations}
\begin{align}
                   g_{8}^{(0)}   = {} & 
                          0 \,,\\ \allowdisplaybreaks[1] 
                   g_{8}^{(1)}   = {} & 
                          0 \,,\\ \allowdisplaybreaks[1] 
                   g_{8}^{(2)}   = {} & 
                          0 \,,\\ \allowdisplaybreaks[1] 
                   g_{8}^{(3)}   = {} & 
                         -4 \,\HPL(0,0,0; x) - 4 \,\zeta_2 \,\HPL(0; x)\,,\\ \allowdisplaybreaks[1] 
                   g_{8}^{(4)}   = {} & 
                         -8 \,\HPL(-1,0,0,0; x)+24 \,\HPL(0,0,-1,0; x)-4 \,\HPL(0,0,0,0; x) \nn \allowdisplaybreaks[1]
                         &-8 \,\HPL(0,0,1,0; x)+8 \,\HPL(0,1,0,0; x)+8 \,\HPL(1,0,0,0; x) \nn \allowdisplaybreaks[1]
                         &-4 \,\zeta_2 (2 \,\HPL(-1,0; x)-3 \,\HPL(0,0; x)-2 \,\HPL(1,0; x)) \nn \allowdisplaybreaks[1]
                         &+4 \,\zeta_3 \,\HPL(0; x)  + 26 \,\zeta _4\,,
\end{align}
\end{subequations}

\begin{subequations}
\begin{align}
                   g_{9}^{(0)}   = {} & 
                          0 \,,\\ \allowdisplaybreaks[1] 
                   g_{9}^{(1)}   = {} & 
                         -\frac{1}{2} \,\HPL(0; x)\,,\\ \allowdisplaybreaks[1] 
                   g_{9}^{(2)}   = {} & 
                         2 \,\HPL(-1,0; x)-\,\HPL(0,0; x) + \zeta_2\,,\\ \allowdisplaybreaks[1] 
                   g_{9}^{(3)}   = {} & 
                         -8 \,\HPL(-1,-1,0; x)+4 \,\HPL(-1,0,0; x)+4 \,\HPL(0,-1,0; x) \nn \allowdisplaybreaks[1]
                         &-2 \,\HPL(0,0,0; x) - 4 \,\zeta_2 \,\HPL(-1; x) + 4 \,\zeta_3\,,\\ \allowdisplaybreaks[1] 
                   g_{9}^{(4)}   = {} & 
                         32 \,\HPL(-1,-1,-1,0; x)-16 \,\HPL(-1,-1,0,0; x)-16 \,\HPL(-1,0,-1,0; x) \nn \allowdisplaybreaks[1]
                         & +8 \,\HPL(-1,0,0,0; x)-16 \,\HPL(0,-1,-1,0; x)+8 \,\HPL(0,-1,0,0; x) \nn \allowdisplaybreaks[1]
                         & +8 \,\HPL(0,0,-1,0; x)-4 \,\HPL(0,0,0,0; x) + 8 \,\zeta_2 (2 \,\HPL(-1,-1; x)\nn \allowdisplaybreaks[1]
                         & -\,\HPL(0,-1; x)) - 4 \,\zeta_3 (4 \,\HPL(-1; x)-\,\HPL(0; x)) + 19 \,\zeta _4\,,
\end{align}
\end{subequations}

\begin{subequations}
\begin{align}
                   g_{10}^{(0)}   = {} & 
                          0 \,,\\ \allowdisplaybreaks[1] 
                   g_{10}^{(1)}   = {} & 
                         \frac{1}{2} \,\HPL(0; x)\,,\\ \allowdisplaybreaks[1] 
                   g_{10}^{(2)}   = {} & 
                         -3 \,\HPL(-1,0; x)+\frac{5}{2} \,\HPL(0,0; x)+\,\HPL(1,0; x) +\zeta_2\,,\\ \allowdisplaybreaks[1] 
                   g_{10}^{(3)}   = {} & 
                         18 \,\HPL(-1,-1,0; x)-14 \,\HPL(-1,0,0; x)-6 \,\HPL(-1,1,0; x)\nn \allowdisplaybreaks[1]
                         & -15 \,\HPL(0,-1,0; x)+\frac{17}{2} \,\HPL(0,0,0; x)+5 \,\HPL(0,1,0; x)\nn \allowdisplaybreaks[1]
                         & -6 \,\HPL(1,-1,0; x)+5 \,\HPL(1,0,0; x)+2 \,\HPL(1,1,0; x) \nn \allowdisplaybreaks[1]
                         & +\frac{1}{2} \,\zeta_2 (-6 \,\HPL(-1; x)+\,\HPL(0; x)-2 \,\HPL(1; x)-6 \log{2}) 
                         -\frac{9 \,\zeta_3}{4}\,,\\ \allowdisplaybreaks[1] 
                   g_{10}^{(4)}   = {} & 
                         -108 \,\HPL(-1,-1,-1,0; x)+80 \,\HPL(-1,-1,0,0; x)\nn \allowdisplaybreaks[1]
                         &+36 \,\HPL(-1,-1,1,0; x)+84 \,\HPL(-1,0,-1,0; x)-44 \,\HPL(-1,0,0,0; x)\nn \allowdisplaybreaks[1]
                         &-28 \,\HPL(-1,0,1,0; x)+36 \,\HPL(-1,1,-1,0; x)-28 \,\HPL(-1,1,0,0; x)\nn \allowdisplaybreaks[1]
                         &-12 \,\HPL(-1,1,1,0; x)+90 \,\HPL(0,-1,-1,0; x)-66 \,\HPL(0,-1,0,0; x)\nn \allowdisplaybreaks[1]
                         &-30 \,\HPL(0,-1,1,0; x)-51 \,\HPL(0,0,-1,0; x)+\frac{41}{2} \,\HPL(0,0,0,0; x)\nn \allowdisplaybreaks[1]
                         &+17 \,\HPL(0,0,1,0; x)-30 \,\HPL(0,1,-1,0; x)+29 \,\HPL(0,1,0,0; x)\nn \allowdisplaybreaks[1]
                         &+10 \,\HPL(0,1,1,0; x)+36 \,\HPL(1,-1,-1,0; x)-28 \,\HPL(1,-1,0,0; x)\nn \allowdisplaybreaks[1]
                         &-12 \,\HPL(1,-1,1,0; x)-30 \,\HPL(1,0,-1,0; x)+17 \,\HPL(1,0,0,0; x)\nn \allowdisplaybreaks[1]
                         &+10 \,\HPL(1,0,1,0; x)-12 \,\HPL(1,1,-1,0; x)+10 \,\HPL(1,1,0,0; x)\nn \allowdisplaybreaks[1]
                         &+4 \,\HPL(1,1,1,0; x)+12 \,\text{Li}_4{\frac12}+\frac{\log^4{2}}{2} 
                         +\frac{1}{2} \,\zeta_2 \left( 24 \log{2} \,\HPL(-1; x) \right. \nn \allowdisplaybreaks[1]
                         &+24 \log{2} \,\HPL(1; x) +12 \,\HPL(-1,-1; x)+4 \,\HPL(-1,0; x)
                         +12 \,\HPL(-1,1; x) \nn \allowdisplaybreaks[1]
                         &-6 \,\HPL(0,-1; x)-11 \,\HPL(0,0; x)-10 \,\HPL(0,1; x)-12 \,\HPL(1,-1; x)\nn \allowdisplaybreaks[1]
                         & + 2 \left.\HPL(1,0; x)-4 \,\HPL(1,1; x)+12 \log^2{2}\right) 
                         + \zeta_3 (20 \,\HPL(-1; x)\nn \allowdisplaybreaks[1]
                         &-14 \,\HPL(0; x)-15 \,\HPL(1; x)) - \frac{95 \,\zeta _4}{8}\,,
\end{align}
\end{subequations}

\begin{subequations}
\begin{align}
                   g_{11}^{(0)}   = {} & 
                          0 \,,\\ \allowdisplaybreaks[1] 
                   g_{11}^{(1)}   = {} & 
                          0 \,,\\ \allowdisplaybreaks[1] 
                   g_{11}^{(2)}   = {} & 
                          0 \,,\\ \allowdisplaybreaks[1] 
                   g_{11}^{(3)}   = {} & 
                         -2 \,\HPL(0,0,0; x) -2 \,\zeta_2 \,\HPL(0; x)\,,\\ \allowdisplaybreaks[1] 
                   g_{11}^{(4)}   = {} & 
                         -4 \,\HPL(-1,0,0,0; x)+4 \,\HPL(0,-1,0,0; x)+12 \,\HPL(0,0,-1,0; x)\nn \allowdisplaybreaks[1]
                         &-6 \,\HPL(0,0,0,0; x)-4 \,\HPL(0,0,1,0; x)+4 \,\HPL(1,0,0,0; x)  \nn \allowdisplaybreaks[1]
                         &-4 \,\zeta_2 (\,\HPL(-1,0; x)-3 \,\HPL(0,-1; x)-\,\HPL(1,0; x)) - \frac{\,\zeta _4}{2}\,,
\end{align}
\end{subequations}

\begin{subequations}
\begin{align}
                   g_{12}^{(0)}   = {} & 
                          0 \,,\\ \allowdisplaybreaks[1] 
                   g_{12}^{(1)}   = {} & 
                          0 \,,\\ \allowdisplaybreaks[1] 
                   g_{12}^{(2)}   = {} & 
                          0 \,,\\ \allowdisplaybreaks[1] 
                   g_{12}^{(3)}   = {} & 
                         -\,\HPL(0,0,0; x) - \zeta_2 \HPL(0; x) \,,\\ \allowdisplaybreaks[1] 
                   g_{12}^{(4)}   = {} & 
                         -2 \,\HPL(-1,0,0,0; x)+2 \,\HPL(0,-1,0,0; x)+2 \,\HPL(0,0,-1,0; x)\nn \allowdisplaybreaks[1]
                         &-3 \,\HPL(0,0,0,0; x)-4 \,\HPL(0,1,0,0; x) + \zeta_2 (-2 \,\HPL(-1,0; x)\nn \allowdisplaybreaks[1]
                         &+6 \,\HPL(0,-1; x)-\,\HPL(0,0; x)) + 2 \,\zeta_3 \,\HPL(0; x) + \frac{\,\zeta _4}{4}\,,
\end{align}
\end{subequations}

\begin{subequations}
\begin{align}
                   g_{13}^{(0)}   = {} & 
                          0 \,,\\ \allowdisplaybreaks[1] 
                   g_{13}^{(1)}   = {} & 
                          0 \,,\\ \allowdisplaybreaks[1] 
                   g_{13}^{(2)}   = {} & 
                         \HPL(0,0; x) + \frac{3 \,\zeta_2}{2}\,,\\ \allowdisplaybreaks[1] 
                   g_{13}^{(3)}   = {} & 
                         -2 \,\HPL(-1,0,0; x)-2 \,\HPL(0,-1,0; x)+4 \,\HPL(0,0,0; x)+4 \,\HPL(1,0,0; x) \nn \allowdisplaybreaks[1]
                         &+\,\zeta_2 (-6 \,\HPL(-1; x)+2 \,\HPL(0; x)-3 \log{2}) - \frac{\,\zeta_3}{4}\,,\\ \allowdisplaybreaks[1] 
                   g_{13}^{(4)}   = {} & 
                         4 \,\HPL(-1,-1,0,0; x)+4 \,\HPL(-1,0,-1,0; x)-8 \,\HPL(-1,0,0,0; x)\nn \allowdisplaybreaks[1]
                         & -8 \,\HPL(-1,1,0,0; x)+4 \,\HPL(0,-1,-1,0; x)-8 \,\HPL(0,-1,0,0; x)\nn \allowdisplaybreaks[1]
                         & -8 \,\HPL(0,0,-1,0; x)+10 \,\HPL(0,0,0,0; x)+12 \,\HPL(0,1,0,0; x)\nn \allowdisplaybreaks[1]
                         & -8 \,\HPL(1,-1,0,0; x)-8 \,\HPL(1,0,-1,0; x)+16 \,\HPL(1,0,0,0; x)\nn \allowdisplaybreaks[1]
                         & +16 \,\HPL(1,1,0,0; x)+12 \,\text{Li}_4{\frac12}+\frac{\log^4{2}}{2} 
                         + 2 \,\zeta_2 \left(12 \log{2} \,\HPL(-1; x) \right. \nn \allowdisplaybreaks[1]
                         & +12 \log{2} \,\HPL(1; x)+6 \,\HPL(-1,-1; x)-2 \,\HPL(-1,0; x)-8 \,\HPL(0,-1; x)\nn \allowdisplaybreaks[1]
                         & \left. +\,\HPL(0,0; x)-12 \,\HPL(1,-1; x)+4 \,\HPL(1,0; x)+3 \log^2{2}\right)  \nn \allowdisplaybreaks[1]
                         & -2 \,\zeta_3 (5 \,\HPL(-1; x)+4 \,\HPL(0; x)+11 \,\HPL(1; x)) -\frac{47 \,\zeta _4}{4}\,,
\end{align}
\end{subequations}

\begin{subequations}
\begin{align}
                   g_{14}^{(0)}   = {} & 
                          0 \,,\\ \allowdisplaybreaks[1] 
                   g_{14}^{(1)}   = {} & 
                          0 \,,\\ \allowdisplaybreaks[1] 
                   g_{14}^{(2)}   = {} & 
                         \HPL(0,0; x)\,,\\ \allowdisplaybreaks[1] 
                   g_{14}^{(3)}   = {} & 
                         -4 \,\HPL(-1,0,0; x)-4 \,\HPL(0,-1,0; x)+5 \,\HPL(0,0,0; x)\nn \allowdisplaybreaks[1]
                         &+2 \,\HPL(0,1,0; x) + \zeta_3\,,\\ \allowdisplaybreaks[1] 
                   g_{14}^{(4)}   = {} & 
                         16 \,\HPL(-1,-1,0,0; x)+16 \,\HPL(-1,0,-1,0; x)-20 \,\HPL(-1,0,0,0; x)\nn \allowdisplaybreaks[1]
                         & -8 \,\HPL(-1,0,1,0; x)+24 \,\HPL(0,-1,-1,0; x)-26 \,\HPL(0,-1,0,0; x)\nn \allowdisplaybreaks[1]
                         & -12 \,\HPL(0,-1,1,0; x)-26 \,\HPL(0,0,-1,0; x)+9 \,\HPL(0,0,0,0; x)\nn \allowdisplaybreaks[1]
                         & +12 \,\HPL(0,0,1,0; x)-12 \,\HPL(0,1,-1,0; x)+8 \,\HPL(0,1,0,0; x)\nn \allowdisplaybreaks[1]
                         & +4 \,\HPL(0,1,1,0; x) - \,\zeta_2 (13 \,\HPL(0,0; x)+2 \,\HPL(0,1; x)) \nn \allowdisplaybreaks[1]
                         & -\zeta_3 (4 \,\HPL(-1; x)+3 \,\HPL(0; x)) - \frac{7 \,\zeta _4}{2}\,,
\end{align}
\end{subequations}

\begin{subequations}
\begin{align}
                   g_{15}^{(0)}   = {} & 
                          0 \,,\\ \allowdisplaybreaks[1] 
                   g_{15}^{(1)}   = {} & 
                          0 \,,\\ \allowdisplaybreaks[1] 
                   g_{15}^{(2)}   = {} & 
                          0 \,,\\ \allowdisplaybreaks[1] 
                   g_{15}^{(3)}   = {} & 
                          0 \,,\\ \allowdisplaybreaks[1] 
                   g_{15}^{(4)}   = {} & 
                         4 \,\HPL(0,-1,0,0; x)-2 \,\HPL(0,0,-1,0; x)-2 \,\HPL(0,1,0,0; x)\nn \allowdisplaybreaks[1]
                         & +4 \,\HPL(1,0,0,0; x) + \zeta_2 (\,\HPL(0,0; x)+4 \,\HPL(1,0; x)) \nn \allowdisplaybreaks[1]
                         & -4 \,\zeta_3 \,\HPL(0; x) + \frac{17 \,\zeta _4}{4}\,,
\end{align}
\end{subequations}

\begin{subequations}
\begin{align}
                   g_{16}^{(0)}   = {} & 
                          0 \,,\\ \allowdisplaybreaks[1] 
                   g_{16}^{(1)}   = {} & 
                          0 \,,\\ \allowdisplaybreaks[1] 
                   g_{16}^{(2)}   = {} & 
                          0 \,,\\ \allowdisplaybreaks[1] 
                   g_{16}^{(3)}   = {} & 
                          0 \,,\\ \allowdisplaybreaks[1] 
                   g_{16}^{(4)}   = {} & 
                         -4 \,\HPL(0,-1,0,0; x)+4 \,\HPL(0,0,-1,0; x)-2 \,\HPL(0,0,0,0; x)\nn \allowdisplaybreaks[1]
                         &-4 \,\HPL(0,0,1,0; x)+4 \,\HPL(0,1,0,0; x)-4 \,\HPL(1,0,0,0; x) \nn \allowdisplaybreaks[1]
                         &-2 \,\zeta_2 (6 \,\HPL(0,-1; x)-\,\HPL(0,0; x)+2 \,\HPL(1,0; x)) - 2 \,\zeta _4\,,
\end{align}
\end{subequations}

\begin{subequations}
\begin{align}
                   g_{17}^{(0)}   = {} & 
                          0 \,,\\ \allowdisplaybreaks[1] 
                   g_{17}^{(1)}   = {} & 
                          0 \,,\\ \allowdisplaybreaks[1] 
                   g_{17}^{(2)}   = {} & 
                          0 \,,\\ \allowdisplaybreaks[1] 
                   g_{17}^{(3)}   = {} & 
                         2 \,(\HPL(0,-1,0; x)-\,\HPL(0,0,0; x)-\,\HPL(0,1,0; x)) 
                         - \zeta_2 \,\HPL(0; x) -\,\zeta_3\,,\\ \allowdisplaybreaks[1] 
                   g_{17}^{(4)}   = {} & 
                         -8 \,\HPL(-1,0,-1,0; x)+8 \,\HPL(-1,0,0,0; x)+8 \,\HPL(-1,0,1,0; x)\nn \allowdisplaybreaks[1]
                         &-20 \,\HPL(0,-1,-1,0; x)+16 \,\HPL(0,-1,0,0; x)+12 \,\HPL(0,-1,1,0; x)\nn \allowdisplaybreaks[1]
                         &+24 \,\HPL(0,0,-1,0; x)-12 \,\HPL(0,0,0,0; x)-16 \,\HPL(0,0,1,0; x)\nn \allowdisplaybreaks[1]
                         &+12 \,\HPL(0,1,-1,0; x)-8 \,\HPL(0,1,0,0; x)-4 \,\HPL(0,1,1,0; x)\nn \allowdisplaybreaks[1]
                         &+8 \,\HPL(1,0,-1,0; x)-8 \,\HPL(1,0,0,0; x)-8 \,\HPL(1,0,1,0; x)  \nn \allowdisplaybreaks[1]
                         &+2 \,\zeta_2 (2 \,\HPL(-1,0; x)+\,\HPL(0,-1; x)+\,\HPL(0,0; x)+\,\HPL(0,1; x)\nn \allowdisplaybreaks[1]
                         &-2 \,\HPL(1,0; x)) + \zeta_3 (4 \,\HPL(-1; x)-\,\HPL(0; x)-4 \,\HPL(1; x)) 
                         -\frac{37 \,\zeta _4}{4}\,,
\end{align}
\end{subequations}

%% file: MIxbox.tex
\section{Master Integrals for the two-loop non-planar Box}
\label{results:xbox2L}
In this Appendix  we present the expression of the $12$
MI's of the  two-loop non-planar Box in Eq.~(\ref{Eq:MIfinBox}).  They are obtained according to the procedure
described in Section~\ref{Sec:xbox}  starting from the integrals $\top{i}$  in Fig.~\ref{Fig:Masters3}.  The latter 
are normalized according to the integration measure (Minkowskian metric is understood)
\begin{displaymath}
\left ( \frac{s^{\eps}\,  \Gamma(1-2\eps)}{\Gamma(1+\eps) \Gamma(1-\eps)^2} \right )^2 \, \int \, \frac{d^D k_1}{\pi^{D/2}} \, \int \,   \frac{d^D k_2 }{\pi^{D/2}}\, .
\end{displaymath}
In the following we collect 
the coefficients $g_{i}^{(a)}$ of the expansion of the MI's around $\eps=0$, 
\begin{displaymath}
g_i = \sum_{a=0}^{4} \eps^a \ g_{i}^{(a)} \ , \qquad i=1,\ldots,12  \, .
\end{displaymath}
The MI's have uniform transcendentality, as can be explicitly checked by using the expressions of the coefficients $g_{i}^{(a)}$.
\begin{subequations}
\begin{align}
                   g_{1}^{(0)} = {} 
                         &-1\,,\\ \allowdisplaybreaks[1] 
                   g_{1}^{(1)} = {} 
                         &-2 \,i\,\pi\,,\\ \allowdisplaybreaks[1] 
                   g_{1}^{(2)} = {} 
                         &12 \,\zeta_2\,,\\ \allowdisplaybreaks[1] 
                   g_{1}^{(3)} = {} 
                         &6 \,\zeta_3 + 8 \,i\,\pi \,\zeta_2\,,\\ \allowdisplaybreaks[1] 
                   g_{1}^{(4)} = {} 
                         &-51 \,\zeta_4 + 12 \,i\,\pi \,\zeta_3\,,
\end{align}
\end{subequations}

\begin{subequations}
\begin{align}
                   g_{2}^{(0)} = {} 
                         &-1\,,\\ \allowdisplaybreaks[1] 
                   g_{2}^{(1)} = {} 
                         &2 \,\HPL(0; x)\,,\\ \allowdisplaybreaks[1] 
                   g_{2}^{(2)} = {} 
                         &-4 \,\HPL(0,0; x)\,,\\ \allowdisplaybreaks[1] 
                   g_{2}^{(3)} = {} 
                         &8 \,\HPL(0,0,0; x) + 6 \,\zeta_3\,,\\ \allowdisplaybreaks[1] 
                   g_{2}^{(4)} = {} 
                         &-16 \,\HPL(0,0,0,0; x) -12 \,\zeta_3 \,\HPL(0; x) + 9 \,\zeta_4\,,
\end{align}
\end{subequations}

\begin{subequations}
\begin{align}
                   g_{3}^{(0)} = {} 
                         &-1\,,\\ \allowdisplaybreaks[1] 
                   g_{3}^{(1)} = {} 
                         &-2 \,\HPL(1; x)\,,\\ \allowdisplaybreaks[1] 
                   g_{3}^{(2)} = {} 
                         &-4 \,\HPL(1,1; x)\,,\\ \allowdisplaybreaks[1] 
                   g_{3}^{(3)} = {} 
                         &-8 \,\HPL(1,1,1; x) + 6 \,\zeta_3\,,\\ \allowdisplaybreaks[1] 
                   g_{3}^{(4)} = {} 
                         &-16 \,\HPL(1,1,1,1; x) +12 \,\zeta_3 \,\HPL(1; x) + 9 \,\zeta_4\,,
\end{align}
\end{subequations}

\begin{subequations}
\begin{align}
                   g_{4}^{(0)} = {} 
                         &\frac{1}{4}\,,\\ \allowdisplaybreaks[1] 
                   g_{4}^{(1)} = {} 
                         &\frac{\,i\,\pi}{2}\,,\\ \allowdisplaybreaks[1] 
                   g_{4}^{(2)} = {} 
                         &-\frac{5 \,\zeta_2}{2}\,,\\ \allowdisplaybreaks[1] 
                   g_{4}^{(3)} = {} 
                         &-\,\zeta_3 - \,i\,\pi \,\zeta_2\,,\\ \allowdisplaybreaks[1] 
                   g_{4}^{(4)} = {} 
                         &-2 \,i\,\pi \,\zeta_3\,,
\end{align}
\end{subequations}

\begin{subequations}
\begin{align}
                   g_{5}^{(0)} = {} 
                         &\frac{9}{4}\,,\\ \allowdisplaybreaks[1] 
                   g_{5}^{(1)} = {} 
                         &-3 \,\HPL(0; x) + \frac{3 \,i\,\pi}{2}\,,\\ \allowdisplaybreaks[1] 
                   g_{5}^{(2)} = {} 
                         &3 \,\HPL(0,0; x) - \frac{15 \,\zeta_2}{2} - 3 \,i\,\pi \,\HPL(0; x)\,,\\ \allowdisplaybreaks[1] 
                   g_{5}^{(3)} = {} 
                         &3 \,\HPL(1,0,0; x) + 15 \,\zeta_2 \,\HPL(0; x) - 15 \,\zeta_3 
                          \nn \allowdisplaybreaks[1]
                         &+ \,i\,\pi (6 \,\HPL(0,0; x)+3 \,\HPL(1,0; x))\,,\\ \allowdisplaybreaks[1] 
                   g_{5}^{(4)} = {} 
                         &-12 \,\HPL(0,0,0,0; x)-6 \,\HPL(0,1,0,0; x)-12 \,\HPL(1,0,0,0; x) \nn \allowdisplaybreaks[1]
                         &-3 \,\HPL(1,1,0,0; x) -15 \,\zeta_2 (2 \,\HPL(0,0; x)+\,\HPL(1,0; x))  \nn \allowdisplaybreaks[1]
                         &+3 \,\zeta_3 (7 \,\HPL(0; x)+\,\HPL(1; x)) - \frac{183 \,\zeta_4}{4}  \nn \allowdisplaybreaks[1]
                         &+i\,\pi \left( -12 \,\HPL(0,0,0; x)-6 \,\HPL(0,1,0; x)-6 \,\HPL(1,0,0; x)\right. \nn \allowdisplaybreaks[1]
                         &-3 \,\HPL(1,1,0; x) -3 \,\zeta_2 \,\HPL(1; x) - 9 \,\zeta_3\left . \right) \,,
\end{align}
\end{subequations}

\begin{subequations}
\begin{align}
                   g_{6}^{(0)} = {} 
                         &\frac{9}{4}\,,\\ \allowdisplaybreaks[1] 
                   g_{6}^{(1)} = {} 
                         &3 \,\HPL(1; x) +\frac{3 \,i\,\pi}{2}\,,\\ \allowdisplaybreaks[1] 
                   g_{6}^{(2)} = {} 
                         &3 \,\HPL(1,1; x) - \frac{15 \,\zeta_2}{2} + 3 \,i\,\pi \,\HPL(1; x)\,,\\ \allowdisplaybreaks[1] 
                   g_{6}^{(3)} = {} 
                         &-3 \,\HPL(0,1,1; x) - 15 \,\zeta_2 \,\HPL(1; x) - 12 \,\zeta_3  \nn \allowdisplaybreaks[1]
                         &+i\,\pi \left( 
                         3 \,\HPL(0,1; x)+6 \,\HPL(1,1; x) - 3 \,\zeta_2  \right) \,,\\ \allowdisplaybreaks[1] 
                   g_{6}^{(4)} = {} 
                         &-3 \,\HPL(0,0,1,1; x)-12 \,\HPL(0,1,1,1; x)-6 \,\HPL(1,0,1,1; x)\nn \allowdisplaybreaks[1]
                         &-12 \,\HPL(1,1,1,1; x) - 15 \,\zeta_2 (\,\HPL(0,1; x)+2 \,\HPL(1,1; x)) \nn \allowdisplaybreaks[1]
                         &-15 \,\zeta_3 \,\HPL(1; x) -\frac{27 \,\zeta_4}{2} 
                         + i\,\pi \left( 3 \,\HPL(0,0,1; x)+6 \,\HPL(0,1,1; x) \right. \nn \allowdisplaybreaks[1]
                         &+6 \,\HPL(1,0,1; x)+12 \,\HPL(1,1,1; x) - 6 \,\zeta_2 \,\HPL(1; x) - 6 \,\zeta_3 \left. \right) \,,
\end{align}
\end{subequations}

\begin{subequations}
\begin{align}
                   g_{7}^{(0)} = {} 
                         & 0 \,,\\ \allowdisplaybreaks[1] 
                   g_{7}^{(1)} = {} 
                         & 0 \,,\\ \allowdisplaybreaks[1] 
                   g_{7}^{(2)} = {} 
                         &\,\HPL(0,0; x) +\,i\,\pi \,\HPL(0; x)\,,\\ \allowdisplaybreaks[1] 
                   g_{7}^{(3)} = {} 
                         &-4 \,\HPL(0,0,0; x)-2 \,\HPL(1,0,0; x) - 6 \,\zeta_2 \,\HPL(0; x)  +2 \,\zeta_3  \nn \allowdisplaybreaks[1]
                         &-i\,\pi \left( 2 (\,\HPL(0,0; x)+\,\HPL(1,0; x)) + 2 \,\zeta_2  \right) \,,\\ \allowdisplaybreaks[1] 
                   g_{7}^{(4)} = {} 
                         &12 \,\HPL(0,0,0,0; x)+4 \,\HPL(0,1,0,0; x)+8 \,\HPL(1,0,0,0; x) \nn \allowdisplaybreaks[1]
                         &+4 \,\HPL(1,1,0,0; x)  
                         +12 \,\zeta_2 (\,\HPL(0,0; x)+\,\HPL(1,0; x)) \nn \allowdisplaybreaks[1]
                         &-4 \,\zeta_3 (\,\HPL(0; x)+\,\HPL(1; x)) +27 \,\zeta_4 \nn \allowdisplaybreaks[1]
                         &+i\,\pi \left( 4 (\,\HPL(0,0,0; x)+\,\HPL(0,1,0; x)+\,\HPL(1,0,0; x)\right.\nn \allowdisplaybreaks[1]
                         &+\,\HPL(1,1,0; x)) + 4 \,\zeta_2 \,\HPL(1; x)\left . \right) \,,
\end{align}
\end{subequations}

\begin{subequations}
\begin{align}
                   g_{8}^{(0)} = {} 
                         & 0 \,,\\ \allowdisplaybreaks[1] 
                   g_{8}^{(1)} = {} 
                         & 0 \,,\\ \allowdisplaybreaks[1] 
                   g_{8}^{(2)} = {} 
                         &\,\HPL(0,0; x)+\,\HPL(0,1; x)+\,\HPL(1,0; x)+\,\HPL(1,1; x) + 3 \,\zeta_2\,,\\ \allowdisplaybreaks[1] 
                   g_{8}^{(3)} = {} 
                         &-4 \,\HPL(0,0,0; x)-2 \,\HPL(0,0,1; x)-2 \,\HPL(0,1,0; x)+2 \,\HPL(1,0,1; x) \nn \allowdisplaybreaks[1]
                         &+2 \,\HPL(1,1,0; x)+4 \,\HPL(1,1,1; x) -6 \,\zeta_2 (\,\HPL(0; x)-\,\HPL(1; x)) + 2 \,\zeta_3\,,\\ \allowdisplaybreaks[1] 
                   g_{8}^{(4)} = {} 
                         &12 \,\HPL(0,0,0,0; x)+4 \,\HPL(0,0,0,1; x)+4 \,\HPL(0,0,1,0; x)\nn \allowdisplaybreaks[1]
                         &-4 \,\HPL(0,1,0,1; x)-4 \,\HPL(0,1,1,0; x)-4 \,\HPL(0,1,1,1; x)\nn \allowdisplaybreaks[1]
                         &-4 \,\HPL(1,0,0,0; x)-4 \,\HPL(1,0,0,1; x)-4 \,\HPL(1,0,1,0; x)\nn \allowdisplaybreaks[1]
                         &+4 \,\HPL(1,1,0,1; x)+4 \,\HPL(1,1,1,0; x)+12 \,\HPL(1,1,1,1; x) \nn \allowdisplaybreaks[1]
                         &+12 \,\zeta_2 (\,\HPL(0,0; x)-\,\HPL(0,1; x)-\,\HPL(1,0; x)+\,\HPL(1,1; x))  \nn \allowdisplaybreaks[1]
                         &-4 \,\zeta_3 (\,\HPL(0; x)-\,\HPL(1; x)) +12 \,\zeta_4\,,
\end{align}
\end{subequations}

\begin{subequations}
\begin{align}
                   g_{9}^{(0)} = {} 
                         & 0 \,,\\ \allowdisplaybreaks[1] 
                   g_{9}^{(1)} = {} 
                         & 0 \,,\\ \allowdisplaybreaks[1] 
                   g_{9}^{(2)} = {} 
                         &\,\HPL(1,1; x) -\,i\,\pi \,\HPL(1; x)\,,\\ \allowdisplaybreaks[1] 
                   g_{9}^{(3)} = {} 
                         &2 \,\HPL(0,1,1; x)+4 \,\HPL(1,1,1; x) + 6 \,\zeta_2 \,\HPL(1; x)  \nn \allowdisplaybreaks[1]
                         &-2 \,i\,\pi (\,\HPL(0,1; x)+\,\HPL(1,1; x))\,,\\ \allowdisplaybreaks[1] 
                   g_{9}^{(4)} = {} 
                         &4 \,\HPL(0,0,1,1; x)+8 \,\HPL(0,1,1,1; x)+4 \,\HPL(1,0,1,1; x)\nn \allowdisplaybreaks[1]
                         &+12 \,\HPL(1,1,1,1; x) +12 \,\zeta_2 (\,\HPL(0,1; x)+\,\HPL(1,1; x)) \nn \allowdisplaybreaks[1]
                         &+i\,\pi \left( -4 (\,\HPL(0,0,1; x)+\,\HPL(0,1,1; x)+\,\HPL(1,0,1; x)\right.\nn \allowdisplaybreaks[1]
                         &+\,\HPL(1,1,1; x)) + 4 \,\zeta_2 \,\HPL(1; x)\left . \right) \,,
\end{align}
\end{subequations}

\begin{subequations}
\begin{align}
                   g_{10}^{(0)} = {} 
                         &-1\,,\\ \allowdisplaybreaks[1] 
                   g_{10}^{(1)} = {} 
                         &-2 \,i\,\pi\,,\\ \allowdisplaybreaks[1] 
                   g_{10}^{(2)} = {} 
                         &17 \,\zeta_2\,,\\ \allowdisplaybreaks[1] 
                   g_{10}^{(3)} = {} 
                         &23 \,\zeta_3 + 18 \,i\,\pi \,\zeta_2\,,\\ \allowdisplaybreaks[1] 
                   g_{10}^{(4)} = {} 
                         &-\frac{317 \,\zeta_4}{2} + 46 \,i\,\pi \,\zeta_3\,,
\end{align}
\end{subequations}

\begin{subequations}
\begin{align}
                   g_{11}^{(0)} = {} 
                         & 0 \,,\\ \allowdisplaybreaks[1] 
                   g_{11}^{(1)} = {} 
                         &\frac{5}{2} (\,\HPL(0; x)+\,\HPL(1; x))\,,\\ \allowdisplaybreaks[1] 
                   g_{11}^{(2)} = {} 
                         &5 \,i\,\pi (\,\HPL(0; x)+\,\HPL(1; x))\,,\\ \allowdisplaybreaks[1] 
                   g_{11}^{(3)} = {} 
                         &-10 \,\HPL(0,0,0; x)-4 \,\HPL(0,0,1; x)-4 \,\HPL(0,1,0; x)-10 \,\HPL(0,1,1; x) \nn \allowdisplaybreaks[1]
                         &-10 \,\HPL(1,0,0; x)-4 \,\HPL(1,0,1; x)-4 \,\HPL(1,1,0; x)-10 \,\HPL(1,1,1; x)  \nn \allowdisplaybreaks[1]
                         &-50 \,\zeta_2 (\,\HPL(0; x)+\HPL(1; x)) + 6 \,\zeta_3 + i\,\pi \left(  
                         -6 (\,\HPL(0,0; x)-\,\HPL(0,1; x) \right. \nn \allowdisplaybreaks[1]
                         &+\,\HPL(1,0; x)-\,\HPL(1,1; x)) - 6 \,\zeta_2\left . \right) \,,\\ \allowdisplaybreaks[1] 
                   g_{11}^{(4)} = {} 
                         &40 \,\HPL(0,0,0,0; x)+16 \,\HPL(0,0,0,1; x)+16 \,\HPL(0,0,1,0; x)\nn \allowdisplaybreaks[1]
                         &-8 \,\HPL(0,0,1,1; x)+8 \,\HPL(0,1,0,0; x)-16 \,\HPL(0,1,0,1; x)\nn \allowdisplaybreaks[1]
                         &-16 \,\HPL(0,1,1,0; x)-40 \,\HPL(0,1,1,1; x)+40 \,\HPL(1,0,0,0; x)\nn \allowdisplaybreaks[1]
                         &+16 \,\HPL(1,0,0,1; x)+16 \,\HPL(1,0,1,0; x)-8 \,\HPL(1,0,1,1; x)\nn \allowdisplaybreaks[1]
                         &+8 \,\HPL(1,1,0,0; x)-16 \,\HPL(1,1,0,1; x)-16 \,\HPL(1,1,1,0; x)\nn \allowdisplaybreaks[1]
                         &-40 \,\HPL(1,1,1,1; x) 
                         + 96 \,\zeta_2 (\,\HPL(0,0; x)-\,\HPL(0,1; x)+\,\HPL(1,0; x)\nn \allowdisplaybreaks[1]
                         &-\,\HPL(1,1; x)) - 55 \,\zeta_3 (\,\HPL(0; x)+\,\HPL(1; x)) + 282 \,\zeta_4 \nn \allowdisplaybreaks[1]
                         &+i\,\pi \left(
                         4 \,\HPL(0,0,0; x)+16 \,\HPL(0,0,1; x)+16 \,\HPL(0,1,0; x) \right.  \nn \allowdisplaybreaks[1]
                         &+4 \,\HPL(0,1,1; x)
                         +4 \,\HPL(1,0,0; x)+16 \,\HPL(1,0,1; x)+16 \,\HPL(1,1,0; x) \nn \allowdisplaybreaks[1]
                         &+4 \,\HPL(1,1,1; x) -36 \,\zeta_2 (\,\HPL(0; x)+\,\HPL(1; x)) + 12 \,\zeta_3\left . \right) \,,
\end{align}
\end{subequations}

\begin{subequations}
\begin{align}
                   g_{12}^{(0)} = {} 
                         &-\frac{1}{4}\,,\\ \allowdisplaybreaks[1] 
                   g_{12}^{(1)} = {} 
                         &\frac{5}{4} \,\HPL(0; x)+\frac{11}{4} \,\HPL(1; x) - 2 \,i\,\pi\,,\\ \allowdisplaybreaks[1] 
                   g_{12}^{(2)} = {} 
                         &-4 \,\HPL(0,0; x)-4 \,\HPL(0,1; x)-4 \,\HPL(1,0; x)-\,\HPL(1,1; x)  +2 \,\zeta_2  \nn \allowdisplaybreaks[1]
                         &+\frac{5}{2} \,i\,\pi (\,\HPL(0; x)+\,\HPL(1; x))\,,\\ \allowdisplaybreaks[1] 
                   g_{12}^{(3)} = {} 
                         &11 \,\HPL(0,0,0; x)+2 \,\HPL(0,0,1; x)+2 \,\HPL(0,1,0; x)-10 \,\HPL(0,1,1; x)\nn \allowdisplaybreaks[1]
                         &+3 \,\HPL(1,0,0; x)-6 \,\HPL(1,0,1; x)-6 \,\HPL(1,1,0; x)-15 \,\HPL(1,1,1; x)  \nn \allowdisplaybreaks[1]
                         &+ \zeta_2 (-13 \,\HPL(0; x)-34 \,\HPL(1; x)) +\frac{9 \,\zeta_3}{2} +i\,\pi \left( 
                         \,\HPL(0,0; x)+4 \,\HPL(0,1; x)\right. \nn \allowdisplaybreaks[1]
                         &+\,\HPL(1,0; x)+7 \,\HPL(1,1; x) + 21 \,\zeta_2\left. \right) \,,\\ \allowdisplaybreaks[1] 
                   g_{12}^{(4)} = {} 
                         &-28 \,\HPL(0,0,0,0; x)+8 \,\HPL(0,0,0,1; x)+8 \,\HPL(0,0,1,0; x)\nn \allowdisplaybreaks[1]
                         &-7 \,\HPL(0,0,1,1; x)-24 \,\HPL(0,1,0,0; x)-24 \,\HPL(0,1,1,1; x)\nn \allowdisplaybreaks[1]
                         &+4 \,\HPL(1,0,0,0; x)+16 \,\HPL(1,0,0,1; x)+16 \,\HPL(1,0,1,0; x)\nn \allowdisplaybreaks[1]
                         &-26 \,\HPL(1,0,1,1; x)-8 \,\HPL(1,1,0,0; x)-8 \,\HPL(1,1,0,1; x)\nn \allowdisplaybreaks[1]
                         &-8 \,\HPL(1,1,1,0; x)-56 \,\HPL(1,1,1,1; x) + \zeta_2 (20 \,\HPL(0,0; x)\nn \allowdisplaybreaks[1]
                         &-31 \,\HPL(0,1; x)+44 \,\HPL(1,0; x)-70 \,\HPL(1,1; x)) \nn \allowdisplaybreaks[1]
                         &+ \frac{1}{2} \,\zeta_3 (-15 \,\HPL(0; x)-11 \,\HPL(1; x)) + \frac{125 \,\zeta_4}{4} \nn \allowdisplaybreaks[1]
                         &+i\,\pi \left( -14 \,\HPL(0,0,0; x)+19 \,\HPL(0,0,1; x)-20 \,\HPL(0,1,0; x)\right. \nn \allowdisplaybreaks[1]
                         &+4 \,\HPL(0,1,1; x)-6 \,\HPL(1,0,0; x)+30 \,\HPL(1,0,1; x)-12 \,\HPL(1,1,0; x)\nn \allowdisplaybreaks[1]
                         &+18 \,\HPL(1,1,1; x) - 6 \,\zeta_2 (9 \,\HPL(0; x) +4 \,\HPL(1; x)) + 78 \,\zeta_3\left . \right) \,,
\end{align}
\end{subequations}